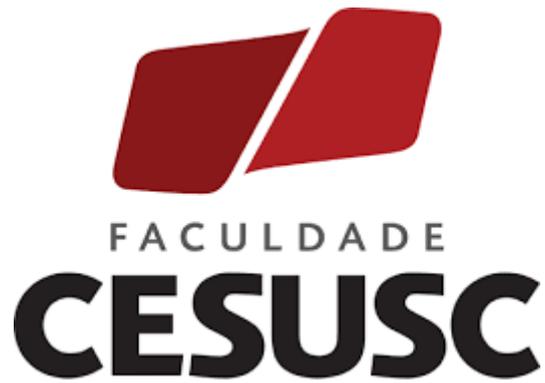

**FACULDADE CESUSC**
**TRABALHO DE CONCLUSÃO DE CURSO**
**CURSO DE DIREITO**

**HEITOR FERREIRA GONZAGA**

**A JURIDICIDADE E A REGULAMENTAÇÃO DOS DARK PATTERNS**

**FLORIANÓPOLIS**

**2022**

**HEITOR FERREIRA GONZAGA**

**A JURIDICIDADE E A REGULAMENTAÇÃO DOS DARK PATTERNS**

Trabalho de Conclusão de Curso apresentado ao Curso de Graduação em Direito da Faculdade CESUSC como requisito à obtenção do título de Bacharel em Direito

Orientador: Paulo Potiara de Alcântara Veloso

**FLORIANÓPOLIS**

**2022**

Este trabalho é dedicado à minha família, amor, e à ciência.

## AGRADECIMENTOS

*"In the night a candle is brighter than the sun"*
(Sting)

# RESUMO

A evolução das interfaces audiovisuais de computadores foi um importante marco para a popularização da internet, sem as quais é impossível conceber o uso desta tecnologia na sociedade moderna. No entanto, o progresso dessas interfaces não tomou caminhos exclusivamente benéficos à humanidade. A partir do início do Século XXI foi constatado acréscimo em padrões de design de interface que, ao contrário de facilitar a navegação, causam danos aos usuários ou restringem suas capacidades decisórias – conferindo-as o nome de *Dark Patterns*, ou Padrões Obscuros. À vista disso, o presente trabalho visa abordar se os *Dark Patterns* são jurídicos ou antijurídicos perante as normas de proteção de dados e de direito do consumidor brasileiros, verificando, ausentes normas específicas sobre a matéria, qual seria a melhor forma de regulá-los. O método de pesquisa empregado é o qualitativo, analisando-se pesquisas, casos judiciais, normas e documentos nacionais e estrangeiros sobre *Dark Patterns*. Após abordar os seus efeitos, seu desenvolvimento legal e estabelecer uma definição compatível com o direito nacional, conclui-se que, apesar de alguns casos serem capazes de produzir danos e violar direitos em alguns casos, a mera declaração da antijuridicidade destas técnicas é uma solução insuficiente, sendo necessárias maiores investigações sobre as hipóteses nas quais os seus impactos negativos são menos aparentes ou quando são utilizados para fins benéficos, entre outras questões. Por isso, sugere-se que a regulamentação dos *Dark Patterns* ocorra através de um sistema composto de leis formais e regulamentos de órgãos da administração pública, através de uma abordagem multidisciplinar e adaptável a novos achados e tecnologias.

Palavras-chave: *Dark Patterns*. Proteção de dados pessoais. Direito do consumidor. Direito digital.

**ABSTRACT**

The evolution of audiovisual computer interfaces was an important milestone for the popularization of the internet, without which it is impossible to conceive the use of this technology in modern society. However, the progress of these interfaces has not taken exclusively beneficial paths for humanity. From the beginning of the 21st century onwards, an increase in interface design patterns was observed that, instead of facilitating navigation, harmed users or restricted their decision-making capabilities – earning them the name of Dark Patterns. In view of this, the present work aims to address whether Dark Patterns are legal or illegal in the face of Brazilian data protection and consumer law, verifying, in the absence of specific norms on Dark Patterns, the best way to regulate them. The research method employed is qualitative, analyzing research, court cases, norms and national and foreign documents on Dark Patterns. After addressing its effects, its legal development and establishing a definition compatible with Brazilian law, it was concluded that, although some implementations are capable of producing damage and violating rights in some cases, the mere declaration of the illegality of these techniques is an insufficient solution, requiring further investigations regarding the hypotheses in which their negative impacts are less apparent or when they are used for beneficial purposes, among other unsolved problems. Therefore, it is suggested that the regulation of Dark Patterns should occur through a system composed of formal laws and regulations of public administration bodies, through a multidisciplinary approach that is adaptable to new findings and technologies.

Keywords: *Dark Patterns*. Data Protection. Consumer Law. Digital Law

# LISTA DE ILUSTRAÇÕES

## LISTA DE ABREVIATURAS E SIGLAS

AADC – Ato do Código de Design Apropriado para Menores da Califórnia.

ANPAD – Associação Nacional de Pós-Graduação e Pesquisa em Administração.

ANPD – Autoridade Nacional de Proteção de Dados.

BEUC – Organização Europeia de Consumidores.

Cal. Civ. Code – Código Civil da Califórnia.

CCPA – Ato sobre a Proteção da Privacidade dos Consumidores da Califórnia.

CDC – Código de Defesa do Consumidor

CPA – Ato sobre Privacidade do Colorado.

CPRA – Ato sobre os Direito à Privacidade dos Consumidores da Califórnia.

DETOUR – Ato para a Redução de Experiências Enganosas a Usuários Online.

DMA – Regulamento Mercados Digitais.

DSA – Regulamento Serviços Digitais.

EDPB – Comitê de Proteção de Dados Europeu.

EFAMRO – Federação Europeia de Associações de Organizações de Pesquisa de Marketing

ESOMAR – Sociedade Europeia de Pesquisa em Opiniões e Marketing

EUA – Estados Unidos da América.

FTC – Comissão Federal de Comércio.

HCI – Interação Humano-Computador.

LGPD – Lei Geral de Proteção de Dados.

LOOs – Grandes Operadores Online.

OCDE – Organização para a Cooperação e Desenvolvimento.

PROCON – Programa de Proteção e Defesa do Consumidor.

RGPD – Regulamento Geral de Proteção de Dados.

ROSCA – Ato para Restaurar a Confiança dos Consumidores Online.

SERNAC – Serviço Nacional do Consumidor.

TSR – Norma Sobre Serviços de Telemarketing.

UI – Interface de Usuário.

UX – Experiência de Usuário.

VCDPA – Ato sobre a Proteção de Dados dos Consumidores da Virgínia.

V.g. – *Verbi Gratia*.

VZBV – Federação de Organizações do Consumidor Alemãs.

# SUMÁRIO

## 1. INTRODUÇÃO

Na lição do jurista espanhol Sebastián Soler, o ordenamento jurídico é construído a partir de uma "rede complexas de bases reais, históricas e culturais"[1] (SOLER, 2019, p. 17; tradução nossa), guardando identidade com as características da população e da sociedade que pretende regular. No entanto, este conjunto complexo de influências acaba por se consolidar no texto legal, uma forma de difícil mutabilidade que atribui ao direito uma pretensão de dever ser autônoma, descompassada com a realidade. Portanto, "a norma é por definição uma cristalização" (SOLER, 2019, p. 77), um recorte temporal, o que não retira de forma alguma a validade do direito, sendo-lhe ínsita essa condição.[2]

O problema da temporalidade só começa a ser verdadeiramente sentido quando as diferenças entre o objeto disposto pela norma jurídica e a realidade social se agravam substancialmente, resultando na perda da eficácia da norma. Por isso o direito se vale dos processos legislativo e judiciário para oxigenar seu conteúdo, as alterando, complementando, interpretando e reinterpretando.

Contudo, o atraso do direito frente ao mundo contemporâneo vem tomando uma preocupante desproporção, demandando-se normas em uma velocidade para a qual o direito não foi pensado, consoante as tecnologias emergentes. As Inteligências Artificiais, por exemplo, são progressivamente mais baratas e mais potentes a cada ano (STANFORD UNIVERSITY, 2022, p. 11), clamando os holofotes do direito especialmente na União Europeia, onde a preocupação com a matéria está gerando uma das primeiras leis para regulá-las (UNIÃO EUROPEIA, 2022).

Outro tópico, de considerável força midiática, é o vício nas redes sociais e suas implicações jurídicas. Reportagem do noticiário Bloomberg (2022) relatou o surgimento de mais de 70 litígios envolvendo o vício dos jovens nessas plataformas, casos para os quais o direito todavia não possui uma solução. No entanto, a despeito do evidente interesse no controle destes efeitos nas redes sociais, que somam

---

[1] "Todo sistema jurídico se asienta sobre una compleja red de bases reales, históricas y culturales [...]"
[2] Segundo Norberto Bobbio, os 3 critérios de valoração da norma jurídica – justiça, validade e eficácia – são independentes entre si, resultando em situações nas quais uma norma jurídica seja justa, mas não válida, e da mesma forma uma norma que seja válida, e não eficaz, entre outras configurações (BOBBIO, 2010)

milhões de usuários todos os dias, essa celeuma possui raízes mais profundas do que inicialmente transparece.

A implementação de interfaces que acarretam variadas consequências negativas aos seus usuários, desde prejuízos psíquicos à constrição do exercício de sua vontade, vem sendo percebida não somente nas redes sociais, mas também em uma pluralidade de websites e aplicativos de celular. O nome convencionado para essas interfaces é *Dark Patterns*, ou padrões obscuros.

*Verbi gratia*, *após* a pressão de inúmeros grupos de defesa do consumidor da União Europeia, a empresa Amazon concordou em simplificar o sistema de cancelamento online do serviço Amazon Prime, que era desproporcionalmente difícil de ser utilizado pelos consumidores. Sobre isso, o Comissário Didier Reynders, do EDPB, foi categórico, afirmando que a inscrição e o cancelamento de um serviço devem ser proporcionais, e por isso "uma coisa é certa: designs manipulativos, ou '*Dark Patterns*', devem ser banidos" (REYNDERS *apud.* UNIÃO EUROPEIA, 2022).

O estudo dos *Dark Patterns* sob a perspectiva do direito é algo muitíssimo recente, de maneira que, apesar de os EUA e a União Europeia já abordarem a temática, não possuímos qualquer legislação brasileira ou movimentos acadêmicos e jurisprudenciais relevantes para contextualizá-los em nosso ordenamento jurídico.

Em virtude disso, este trabalho buscou compreender qual é a juridicidade dos *Dark Patterns* no direito nacional e a melhor forma de regulá-los, partindo-se da hipótese que, considerados seus efeitos potencialmente degradantes aos usuários da internet, eles sejam antijurídicos e devem ser proibidos. Delimitando o escopo da pesquisa, os *Dark Patterns* foram apreciados segundo a ótica dos direitos do consumidor e da proteção de dados pessoais, áreas escolhidas por guardarem maior afinidade com o tema estudado.

À conclusão, no entanto, a hipótese do trabalho foi parcialmente refutada. Ao longo da pesquisa, os *Dark Patterns* se mostraram como objeto de estudo mais complexo que o antevisto, possuindo nuances que atrapalham o enquadramento de todas as interfaces no modelo binário jurídico-antijurídico, em razão do contexto em que são aplicadas e a necessidade de maiores esclarecimentos sobre as suas influências. Assim, enquanto existem casos nos quais uma particular interface substantivamente causa danos e viola direitos, estes resultados não se repetem todas as vezes em que ela é empregada, e da mesma forma para outras interfaces classificadas como *Dark Patterns* pela doutrina.

Neste sentido, a proibição derradeira dessas técnicas tampouco é a melhor solução, a partir do que sugeriu-se um sistema regulatório em que participem conjuntamente, além do Poder Legislativo, órgãos especiais da administração pública e entidades setoriais para estabelecer normas, diretrizes e orientações específicas e objetivas sobre os *Dark Patterns*, bem como para incentivar e coordenar a pesquisa sobre eles, desenvolvendo uma "malha" normativa capaz de equacionar a densidade da matéria, os diferenciados contextos de sua utilização e as mudanças decorrentes do surgimento de novas tecnologias.

A pesquisa foi dividida em 3 partes. A primeira se debruçou sobre os contornos iniciais dos *Dark Patterns*, trazendo o seu contexto histórico, seus mecanismos de funcionamento, sua eficácia e uma amostra da doutrina proeminente sobre a matéria, com o intuito de destacar os elementos de pertinência jurídica a serem considerados. A segunda visou estudar como os *Dark Patterns* estão sendo regulados pelo direito estrangeiro, estabelecendo critérios comparativos para analisar essas técnicas segundo o direito brasileiro. Por fim, a terceira atacou o cerne da questão, analisando sua juridicidade e como regulamentá-los, o que foi feito através de quatro subtópicos: (i) como os *Dark Patterns* vêm sendo explorados academicamente no país; (ii) como conceituá-los no direito brasileiro; (iii) de que maneira eles interagem com as normas de direito do consumidor e de proteção de dados pessoais; e (iv) qual estratégia de regulamentação é adequada para discipliná-los.

## 2. APRESENTAÇÃO, DEFINIÇÃO E EFEITOS DOS *DARK PATTERNS*

Inicialmente desenhada pela "Agência Avançada de Projetos de Pesquisa" (*Advanced Research Projects Agency*, ou ARPA; tradução nossa), a rede descentralizada de computadores ARPANET foi desenvolvida com o intuito de permitir o compartilhamento de recursos especializados (hardware, software, serviços e aplicações) entre pesquisadores (COHEN-ALMAGOR, 2011, p. 47), e possivelmente para prevenir que os seus sistemas não fossem destruídos num só bombardeio, consideradas as tensões da Guerra Fria (ANDRADE; MAGRO, 2021, p. 21)[3]. A princípio, o interesse na internet era exclusivamente acadêmico/institucional.

"Navegar na internet" só se tornou acessível à população leiga depois de árduos e insistentes esforços para o desenvolvimento de aplicações amigáveis -

---

[3] Nas palavras de Américo Ribeiro Magro e Landolfo Andrade sobre a criação da internet, "criada no auge da Guerra Fria, esta rede de comunicações foi concebida - para além de sua evidente celeridade na transmissão de informações - de modo a resistir à eventualidade de um ataque nuclear da ex-União Soviética que incapacitasse, total ou parcialmente, a habilidade de transmissão de ordens do governo dos Estados Unidos, sendo, portanto, baseada em um sistema descentralizado" (ANDRADE; MAGRO, 2021, p. 21) A intenção de prevenir ataques que desabilitassem a rede, no entanto, não é pacificamente aceita pelos historiadores. Sobre isto, Raphael Cohen-Almagor afirma: "Um mito popular afirma que os cientistas do Departamento de Defesa pensavam que, se os soviéticos fossem capazes de lançar satélites, eles poderiam também ser capazes de lançar mísseis nucleares de longa distância. Como as redes na época dependiam de uma única função de controle central, diz o mito, a principal preocupação era a vulnerabilidade das redes a ataques: uma vez que o ponto de controle central da rede deixasse de funcionar, toda a rede se tornaria inutilizável. Os cientistas queriam difundir a rede para que pudesse ser sustentada após o ataque a um ou mais de seus centros de comunicação (Schneider & Evans, 2007). Eles tinham em mente um "repositório descentralizado para segredos relacionados à defesa" durante a guerra (Conn, 2002). , p. xiii). No entanto, os pioneiros do projeto ARPA Network argumentam que a ARPANET não estava relacionada à construção de uma rede resistente à guerra nuclear: "Isso nunca foi verdade para a ARPANET, apenas o estudo RAND não relacionado sobre voz segura considerou guerra nuclear. No entanto, o trabalho posterior sobre Internet enfatizou a robustez e a capacidade de sobrevivência, incluindo a capacidade de suportar perdas de grandes porções das redes subjacentes. Leonard Kleinrock, o pai da *Modern Data Networking*, um dos pioneiros da comunicação em rede digital que ajudou a construir a ARPANET, explicou que a razão pela qual a ARPA queria implantar uma rede era permitir que seus pesquisadores compartilhassem recursos especializados uns dos outros (hardware, software, serviços e aplicativos). Não era para se proteger contra um ataque militar. E David D. Clark, pesquisador sênior do Laboratório de Ciência da Computação do MIT que trabalhou no projeto ARPANET no início dos anos 1970, disse que nunca ouviu falar em capacidade de sobrevivência nuclear e que não há menção dessa ideia nos registros da ARPA da década de 1960. Em uma comunicação pessoal, Clark escreveu: Perguntei a algumas das pessoas que pressionaram pela ARPAnet: Larry Roberts e Bob Kahn. Ambos afirmam que ninguém tinha a capacidade de sobrevivência nuclear em mente. Eu estive lá por volta de 73, e nunca ouvi isso uma vez. Pode ter havido alguém que teve a ideia no fundo de sua mente, mas 1) se sim, ele manteve-a para si e 2) não consigo descobrir quem poderia ter sido. Sabemos quem eram mais ou menos todos os atores importantes. (Infelizmente, Licklider morreu, mas acho que perguntei a ele quando ele ainda estava vivo. Gostaria de ter notas melhores.) Portanto, estou muito confiante de que o objetivo de Baran não sobreviveu para impulsionar o esforço da ARPA. Era compartilhamento de recursos, interação humana... e comando e controle."" (COHEN-ALMAGOR, 2011, p. 47; tradução nossa)

capazes de sintetizar a complexa carga de informações disponível na rede em poucos botões e símbolos amplamente reconhecíveis. *Exempli gratia*, a popularização do uso da internet efetivamente tomou forças com o lançamento do navegador web "Mosaic", desenvolvido em 1993 por especialistas do "Centro Nacional para Aplicações de Supercomputação" (*National Center for Supercomputing Applications*, ou NCSA; tradução nossa), que continha elementos de interface gráfica intuitivos, práticos e muitos superiores à concorrência (MAGRO; ANDRADE, 2021, p. 27). A descrição de Glenn Ricart sobre o Mosaic, em exemplar de 1994 da revista *Computer in Physics*, mostra quão deslumbrante foi a introdução desse software à época:

> Há uma empolgação em usar o MOSAIC. É como a aventura de caminhar pelos corredores de uma livraria recém-descoberta, encontrar uma nova exposição na galeria de arte ou museu, ou deixar-se atrair pelo fascínio de ver e ouvir as lojas de um novo shopping de luxo. A aventura parece semelhante porque esses são exatamente os tipos de coisas pelas quais você pode navegar o quanto quiser com o navegador de Internet multimídia MOSAIC. A experiência é renovada diariamente à medida que milhares de organizações em todo o mundo competem por sua atenção com as novas ofertas do MOSAIC.
> [...]
> Assim como o GOPHER da Internet, o MOSAIC também é um navegador, mas possui recursos de multimídia integrados. MOSAIC é um leitor de hipertexto com links que podem percorrer a Internet em todo o mundo. Se você já tentou usar a Internet, mas achou difícil navegar, o MOSAIC vai te surpreender. O MOSAIC requer apenas um controle firme do mouse de seu Macintosh, um PC executando o Microsoft Windows ou uma estação de trabalho Unix com X-windows. (RICART, p. 249; tradução nossa)[4]

Modernamente, a criação de interfaces pensando no engajamento humano com sistemas de computador é feita por profissionais de HCI e UX/UI. HCI/CHI ou IHC (*Human-Computer Interaction/Computer-Human Interaction* ou, em vernáculo, Interação Humano-Computador) é uma área de pesquisa multidisciplinar que aplica elementos de psicologia, antropologia, design gráfico e outras matérias para, a partir do método científico, solucionar problemas de design (MACKAY; FAYARD, 1997).

---

[4] "There is an excitement that comes with using MOSAIC. It is like the adventure of walking down the aisles of a newly discovered bookstore, or finding a fresh exhibit at the art gallery or museum, or allowing yourself to be attracted by the come-hither allure of sight and sound from the stores in a new upscale mall. The adventure seems similar because these are the very kinds of things you can browse to your heart 's content with the MOSAIC multimedia Internet browser. The experience is refreshed daily as a thousand organizations across the globe vie for your attention with new MOSAIC offerings. [...] Like the Internet GOPHER, MOSAIC is also a browser, but it has built in multimedia capabilities. MOSAIC is a hypertext reader with links that can roam around the world-wide Internet. If you have tried to use the Internet but found it difficult to navigate , MOSAIC will amaze you. MOSAIC requires only a firm grip on the mouse of your Macintosh, a PC running Microsoft Windows, or a Unix workstation with X-windows".

UX/UI é um ramo de desenvolvimento de HCI.[5] UX significa Experiência de Usuários (*User Experience;* tradução nossa) e "se refere à experiência geral relacionada com a percepção (emoções e pensamentos), a reação e o comportamento que um usuário sente e pensa através do uso direto ou indireto de um sistema, produto, conteúdo ou serviço" (JOO, 2017, p. 9931; tradução nossa)[6]. UI significa Interface de Usuário (*User Interface*; tradução nossa), se referindo "a um sistema e a um usuário interagindo entre si por meio de comandos ou técnicas para operar o sistema, inserindo dados e usando conteúdos" (JOO, 2017, p. 9931; tradução nossa)[7]. O objetivo da UX/UI, desta forma, é desenvolver interfaces amigáveis que proporcionem a melhor experiência possível aos seus usuários.

Em observância à essencialidade desses ofícios para a lisa interação do ser humano com o ciberespaço, assim como aos potenciais malefícios que poderiam advir se conduzidos incorretamente, a responsabilidade para com a criação de ferramentas de design é recorrente alvo de estudos pelos profissionais de UX/UI. Gray *et al.* (2018, p. 2-3) pontuam que o desenvolvimento de uma interface deve levar em conta fatores éticos, a partir do uso de "métodos valorativamente sensitivos, design crítico-reflexivo e design persuasivo" (tradução nossa)[8]. Bösch, *et al.* (2016, p. 240), com enfoque na proteção da privacidade, elencam sete princípios que devem ser adotados no desenvolvimento de padrões de design[9]: "proativo e não reativo; privacidade como configuração primária; privacidade como parte do design; funcionalidade plena; segurança *end-to-end* (do começo ao fim do tratamento de dados); visibilidade e transparência; e respeito à privacidade dos usuários" (tradução nossa)[10].

---

[5] UX/UI difere de HCI ao ser orientada à construção de interfaces gráficas. O HCI, enquanto área de pesquisa abrangente, reunindo elementos de ciências sociais, exatas e design e, por isso, não é orientado concretamente ao design ou à ciência, mas a uma mistura indiferenciada de ambos para solucionar problemas específicos (MACKAY; FAYARD, 1997, p. 226).

[6] "User Experience (UX) refers to the overall experience related to the perception (emotion and thought), reaction, and behavior that a user feels and thinks through his or her direct or indirect use of a system, product, content, or service".

[7] "A user interface (UI) refers to a system and a user interacting with each other through commands or techniques to operate the system, input data, and use the contents".

[8] "Value-Sensitive Methods, Critical and Reflective Design e Persuasive Design".

[9] "A ideia de um padrão é capturar a uma instância de um problema e sua correspondente solução, abstraí-la de um caso de uso específico, e moldá-lo de maneira mais genérica, para que possa ser aplicado e reutilizado em diferentes cenários compatíveis." (BÖSCH et al, 2018; tradução nossa).

[10] "Proactive not reactive; privacy as the default setting; privacy embbed to design; full functionality; end-to-end security; visibility and transparency; respect for user privacy".

Contudo, estas diretrizes são meramente teóricas, não impedindo que empresas e profissionais de UX/UI criem designs antiéticos, desenvolvendo ferramentas para confundir os usuários, dificultar a tomada de certas decisões, manipulá-los a praticar ações inconsistentes com suas preferências (LUGURI; STRAHILEVITZ, 2021, p. 44) e, inclusive, capturar indevidamente os seus dados pessoais (GRAY *et al.*, 2018, p. 8). O nome conferido a essas técnicas de design, apenas recentemente estudadas pela doutrina estrangeira e praticamente desconhecidas no Brasil, é *Dark Patterns*, ou Padrões Escuros[11], na língua vernácula.

Um exemplo claro do uso de *Dark Patterns* é o empregado pela empresa IObit, no seu software assistente de instalação do programa Driver Booster 10.

Figura 1 - Assistente de instalação do Driver Booster 10

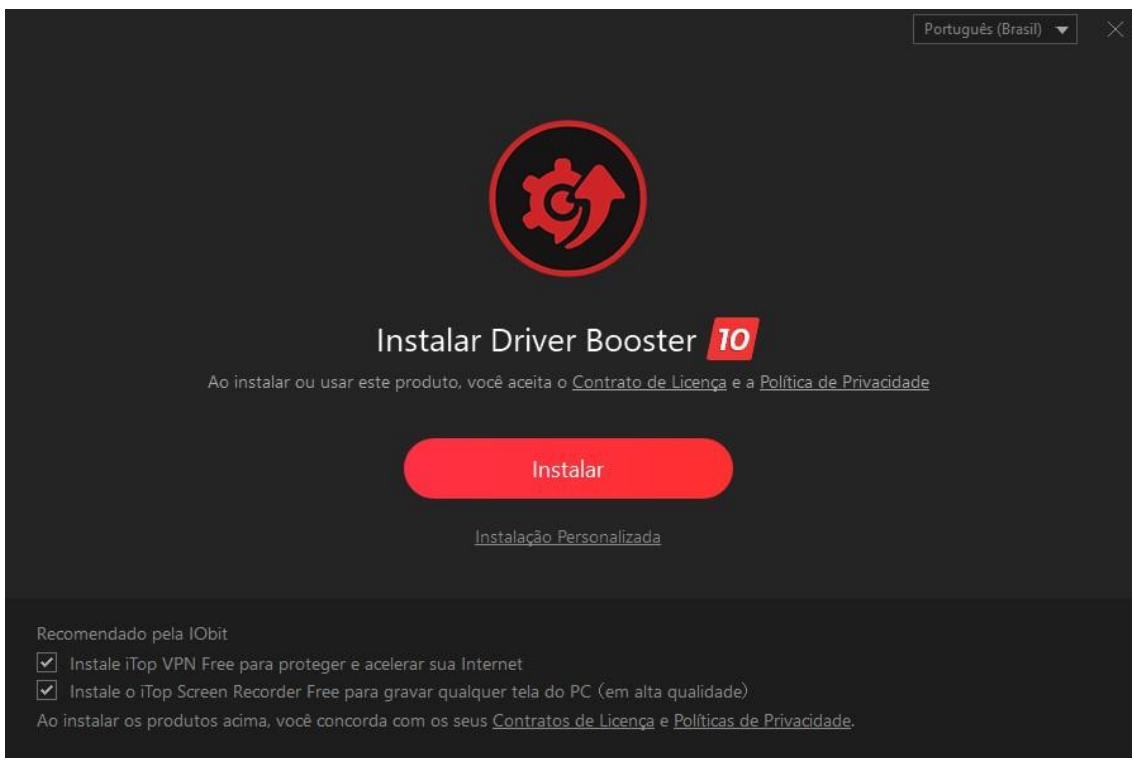

Fonte: criado pelo autor.

---

[11] Vale ressaltar que o uso do termo *Dark Patterns* pela doutrina se popularizou recentemente, sendo possível citar a atribuição de outros nomes para designar este tipo de técnica, como *evil interfaces* ou *malicious interfaces* ("interfaces maliciosas", em tradução livre), segundo Gregory Conti e Edward Sobiesk (2009), utilizados em um trabalho sobre o uso antiético de interfaces, seus efeitos e contramedidas. No Brasil, o conceito "interface maliciosa" foi utilizado por Lemos e Marques em 2019, sendo proposto como alternativa ao estrangeirismo "*Darks Patterns*". Contudo, a despeito do termo sugerido em língua portuguesa, para os fins deste trabalho o conceito será mantido em inglês - *Dark Patterns* -, em razão de ter sido a nomenclatura proeminente na doutrina especializada, fomentando a precisão e unidade conceitual do objeto de estudo nos escopos de pesquisa nacional e internacional.

Observando os elementos da figura 1, a interface do assistente de instalação do software é imbuída do título "Instalar Driver Booster 10" e de um chamativo botão vermelho com a palavra "instalar", levando um usuário desatento a clicá-lo sem levar em consideração o restante das informações, apresentadas em fonte e cor de tal forma a passarem facilmente despercebidas. Isto é feito de forma proposital para evitar que se leia os textos, pois a mensagem imediatamente abaixo do título se refere ao contrato de licença e à política de privacidade da aplicação, enquanto os demais, situados na margem inferior da tela, são autorizações pré-selecionadas para a instalação de outros aplicativos da IObit. Os *Dark Patterns* utilizados foram destacados em amarelo na imagem abaixo.

Figura 2 - Dark Patterns no assistente de instalação do Driver Booster 10.

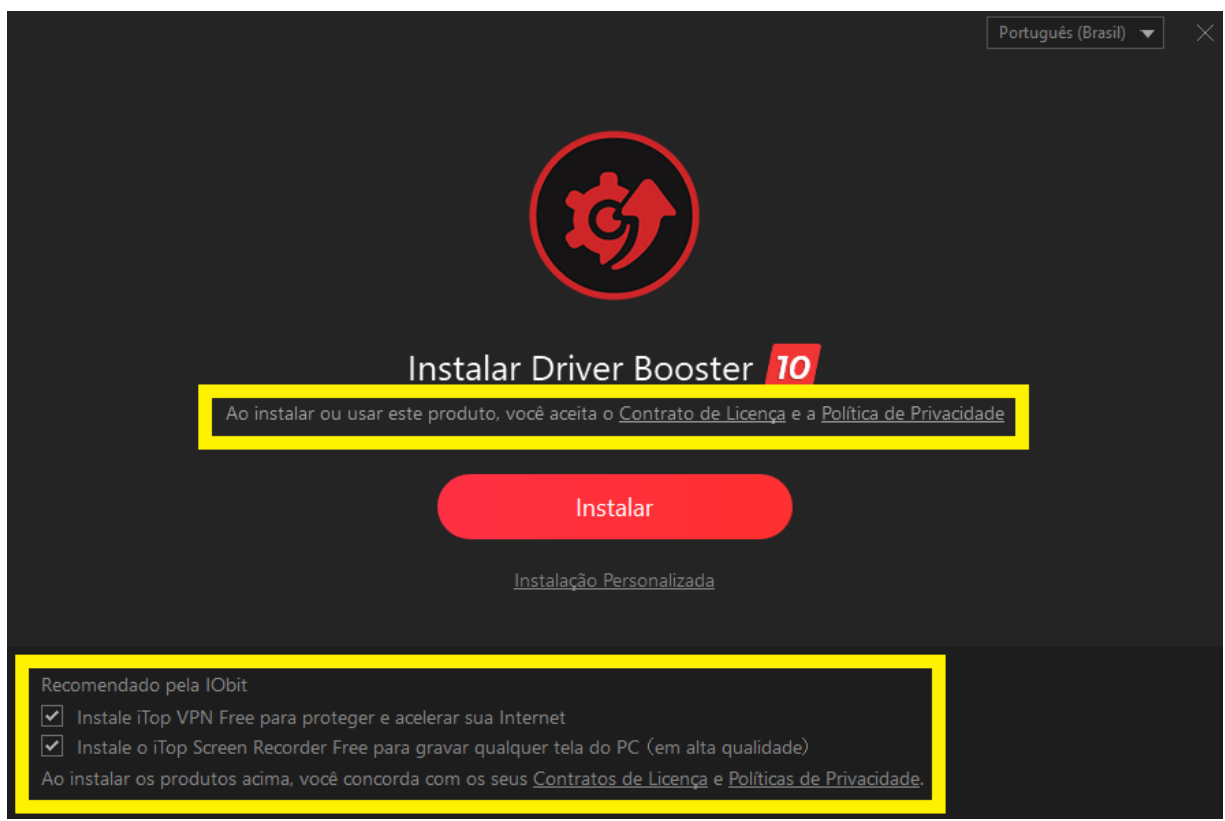

Fonte: criado pelo autor.

Em suma, a interface do referido software é arquitetada de maneira que um usuário comum pense estar simplesmente instalando o Driver Booster 10 quando, na realidade, foi enganado a adicionalmente baixar os aplicativos iTop VPN Free e o iTop Screen Recorder, assim como também foi direcionado para longe dos termos do contrato e da política de privacidade do aplicativo, por conta do design de pouco contraste usado na mensagem.

Estas técnicas rapidamente se alastraram pela internet em uma pluralidade de tipos e funções, algumas mais evidentes e outras mais sutis. Buscando averiguar a frequência do uso desses Padrões e a quantidade de suas espécies, o Serviço Nacional do Consumidor do Chile (*Servicio Nacional del Consumidor*, ou SERNAC; tradução nossa) realizou levantamento quantitativo automatizado nos websites de 107 empresas que participaram do Cybermonday 2020 (SERNAC, 2021, p. 7), um dos eventos sobre compras online mais relevantes do país (CYBER.CL, s.d.).

O rol de classificações de *Dark Patterns* do relatório do SERNAC teve como base a pesquisa de UX Harry Brignull, através da organização sem fins lucrativos de endereço eletrônico "www.darkpatterns.org", e o estudo "*Dark Patterns* em escala: achados de um *Crawl* em 11 mil websites de compras" (tradução nossa)[12] (SERNAC, 2021, p. 8).

Entre os mais prevalentes, foram destacados os *Dark Patterns* das espécies "sinais de urgência e escassez" (*Señales de urgencia y escasez*), que incitam o usuário a tomar uma ação com base em algum indicativo de iminente esgotamento da oportunidade; notificações de atividade (*notificaciones de actividad*), o recebimento frequente de notificações de atividades de outros usuários para estimular o engajamento com a plataforma; "hotéis de baratas" (*Roach Motels*), situações nas quais os usuários entram com facilidade, mas que depois se verificam muito difíceis de sair, como processos de cancelamento de serviços projetados para serem demasiadamente longos e cansativos; e adições automáticas aos carrinhos de compras (*Colarse en el carrito de compra*), quando o website acrescenta inadvertidamente ao seu carrinho um item relacionado às suas pesquisas na plataforma. (SERNAC, 2021, p. 8-9; tradução nossa)

O estudo concluiu que 64% das empresas participantes do evento usaram *Dark Patterns* para captar usuários através de suas plataformas. Os mais incorporados foram os dos tipos "ação forçada" (*acciones forzadas*), que "obrigam o usuário a tomar alguma ação para completar sua compra, como a inscrição em uma *newsletter*" (tradução nossa)[13] (30 empresas), "sinais de urgência e escassez" (25

---

[12] "Dark Patterns at Scale: Finding from a Crawl of 11K Shopping Websites"
[13] "Obligan al usuario a hacer algo para completar su compra. Por ejemplo: registrar una cuenta o suscribirse a un boletín informativo"

empresas), e "testemunhos duvidosos" (*testimonios dudosos*), "testemunhos de clientes cuja origem não está clara" (SERNAC, 2021, p. 9; tradução nossa)[14] - (18 empresas) (SERNAC, 2021, p. 8-16).

Outra pesquisa mostrou resultados ainda mais significativos. Mathur *et al.* (2019) construíram um *web crawler*[15] para, com o auxílio de outras técnicas de ciência da computação, procurar por *Dark Patterns* nos 11.000 mais frequentados comércios eletrônicos da internet, conforme classificação do sistema de ranqueamento *Alexa Top Sites*, a partir do que criaram um esquema taxonômico para classificar os padrões encontrados. (MATHUR *et al.*, 2019, p. 3)[16]

Do total de sites analisados, os pesquisadores identificaram 1.818 *Dark Patterns* distribuídos entre 1.254 sites, somando aproximadamente 11.1% do total de alvos analisados (MATHUR *et al.*, 2019, p. 3). Não fosse o bastante, os resultados não correspondem com a exata quantidade de *Dark Patterns* nessas plataformas, pois o *web crawler* era somente capaz de vascular textos, ou seja, era incapaz de descobrir Padrões inseridos dentro de imagens (MATHUR *et al.*, 2019, p. 3). O estudo também revelou que os comércios eletrônicos que mais utilizaram essas técnicas são também alguns dos melhor pontuados no ranqueamento de popularidade Alexa. (MATHUR *et al*, 2019, p. 3). Os *Dark Patterns* descobertos foram identificados e classificados em 7 categorias e 15 espécies, usando as nomenclaturas cunhadas por Brignull em 2010 (MATHUR *et al.*, 2019, p.13).

---

[14] "Testimonios de clientes cuyo origen no está claro" (SERNAC, 2021, p. 9)

[15] "A WWW nos fornece grandes quantidades de informações úteis disponíveis eletronicamente como hipertexto. Este grande conjunto de hipertextos está mudando de forma dinamicamente e semanticamente desestruturada, tornando difícil encontrarmos informações valiosas. Portanto, um *web crawler* para a descoberta automática de informações valiosas da *Web*, ou *Web Mining*, é importante para nós nos dias de hoje. Na realidade, um *web crawler* é um programa que percorre automaticamente a *web* baixando documentos e seguindo links de página em página. Eles são usados principalmente por motores de pesquisa para coletar dados para indexação. Outras aplicações possíveis incluem a validação de páginas, análise estrutural e visualização, notificações de atualização, espelhamento e assistentes/agentes pessoais da web etc. Os *web crawlers* também são conhecidos como aranhas, robôs, vermes etc." (DHENAKARAN; SAMBANTHAN, 2011, p. 265; tradução nossa)

[16] "Neste artigo, apresentamos uma abordagem automatizada que permite que especialistas identifiquem *Dark Patterns* em escala na web. Nossa abordagem se baseia em (1) um rastreador da web, construído sobre OpenWPM – uma plataforma de medição de privacidade na web – para simular uma experiência de navegação do usuário e identificar elementos da interface do usuário; (2) um agrupamento de texto para extrair todos os designs de interface do usuário dos dados resultantes; e (3) em inspecionar os aglomerados resultantes em busca de instâncias de *Dark Patterns*. Também desenvolvemos uma taxonomia para que os pesquisadores possam compartilhar terminologia descritiva e comparativa para explicar como os *Dark Patterns* subvertem a tomada de decisão do usuário e levam a danos. Baseamos essa taxonomia nas características dos padrões escuros, bem como nos vieses cognitivos que eles exploram nos usuários" (MATHUR et al, 2019, p. 3; tradução nossa)

Estes resultados, inclusive o relatório do SERNAC, são algumas das poucas tentativas de enumerar e classificar os *Dark Patterns* para fins jurídicos. Em contrapartida, no âmbito do design de UX/UI a discussão ética sobre o uso de *Dark Patterns* já foi discutida com bastante rigor: trabalhos como os produzidos por Zagal, Björk e Lewis (2013), e Gray *et al.* (2018) traçam diretrizes sólidas sobre a aplicação destas interfaces, abordando categorias e seus efeitos em diferentes searas. Os três primeiros (ZAGAL *et al*, 2013, p. 1-8) elaboraram um estudo classificatório sobre os *Dark Patterns* aplicados a jogos de videogame, estabelecendo a proto-definição de "*Dark Pattern* de design de jogos" como "um padrão usado intencionalmente pelos criadores do jogo para causar experiências negativas sem o consentimento dos jogadores" (tradução nossa)[17]. Gray *et al.* (2018) parte de uma perspectiva mais abrangente sobre *Dark Patterns*, os categorizando e abordando segundo os princípios do design ético.

No entanto, sem menosprezo aos esforços acadêmicos despendidos até o momento, as características que definem um *Dark Pattern* foram trabalhadas pela doutrina de forma pouco uniforme, contraditória ou equivocada. Nesse caminho, as taxonomias propostas para nomear cada um de seus tipos são em grande parte incompletas e consideravelmente díspares. Como salienta Mathur *et al.* (2021) acerca dos resultados de seu estudo sobre as definições de *Dark Patterns* utilizadas na literatura:

> [...] há significativa variância entre as facetas [dos *Dark Patterns*] refletidas em definições. Por exemplo, nove definições não envolvem qualquer característica de uma interface de usuário, quatro definições não especificam o mecanismo de efeito sobre o usuário, oito definições não contemplam os papéis dos designers de interface, e dez definições não envolvem elementos de benefício ou prejuízo. [...] Também há significativa variância entre cada faceta das definições. Por exemplo, Brignull e Conti e Sobiesk introduzem definições que envolvem a característica da interface de usuário, mas aquele descreve os *Dark Patterns* como "truques", enquanto estes afirmam que *Dark Patterns* são "maliciosos".[18] (MATHUR *et al.*, 2021, p. 4-5; tradução nossa)

---

[17] "Proto-definition 2: A dark game design pattern is a pattern used intentionally by a game creator to cause negative experiences for players without those players' consent."

[18] "As the table shows, there is significant variation among the facets reflected in definitions. For instance, nine definitions do not involve any characteristic of the user interface, four definitions do not specify a mechanism of effect on users, eight definitions do not address the role of user interface designers, and ten definitions do not involve benefit or harm elements. There is also significant variation within each facet of the definitions. For example, Brignull and Conti and Sobiesk both have definitions that involve characteristics of user interfaces, but the former describes dark patterns as "tricks" while the latter notes that dark patterns are "malicious.""

Visando assegurar uma maior precisão conceitual sobre estas técnicas, Mathur *et al.* (2021) lista uma série de atributos inerentes a determinados tipos de *Dark Patterns*. Estes atributos são:

a) "assimétricos": "[...] impõe ônus desiguais sobre as escolhas disponíveis ao usuário. As escolhas que beneficiam o serviço são destacadas proeminentemente, enquanto as que beneficiam o usuário são tipicamente escondidas [...] ou tiradas fora de vista [...]"[19] (tradução nossa) (MATHUR *et al.,* 2021, p. 5).

b) "disfarçados": "*Dark Patterns* disfarçados empurram o usuário em direção a selecionar determinadas decisões ou resultados, mas escondem do usuário a influência desse mecanismo."[20] (MATHUR *et al.*, 2021, p. 8; tradução nossa)

c) Enganosos". "*Dark Patterns* enganosos induzem falsas crenças em usuários a partir de afirmações distorcidas ou enganosas ou de omissões."[21] (MATHUR *et al.*, 2021, p. 8; tradução nossa)

d) "Que escondem informações": "*Dark Patterns* que escondem informações obscurecem ou atrasam a apresentação de informações necessárias aos usuários."[22] (MATHUR *et al.*, 2021, p. 8; tradução nossa)

e) "restritivos": "*Dark Patterns* restritivos reduzem ou eliminam as escolhas apresentadas aos usuários."[23] (MATHUR *et al.*, 2021, p. 8; tradução nossa)

f) "De tratamento diferenciado": *Dark Patterns* que "[...] tratam um grupo de pessoas de maneira diferente e desvantajosa em comparação a outro grupo. [...]"[24] (MATHUR *et al.*, 2021, p. 8; tradução nossa).

Estas características são fruto do arranjo bibliográfico e empírico organizado pelos autores, com base nos resultados da pesquisa com o *web crawler* em 2019, proporcionando alguns fundamentos para entender os diversos tipos de *Dark Patterns* na internet. No entanto, advindo da análise de um número todavia limitado de pesquisas, o rol levantado não deve ser tido como absoluto, pendendo mais estudos a respeito.

---

[19] "Asymmetric dark patterns impose unequal burdens on the choices available to the user. The choices that benefit the service are feature prominently while the options that benefit the user are typically tucked away behind several clicks or are obscured from view by varying the style and position of the choice."
[20] "Covert dark patterns push a user toward selecting certain decisions or outcomes, but hide the influence mechanism from the user"
[21] "Deceptive dark patterns induce false beliefs in users through affirmative misstatements, misleading statements, or omissions."
[22] "Information hiding dark patterns obscure or delay the presentation of necessary information to users."
[23] "Restrictive dark patterns reduce or eliminate the choices presented to users."
[24] "[...] treat one group of users differently from another. [...] for example, the Pay to Skip interface on gaming sites allows users with more resources to gain an edge over users who cannot afford to pay."

Esta "primeira onda" tipológica (LUGURI; STRAHILEVITZ, 2021, p. 44) produziu resultados que podem ser resumidos na seguinte tabela de classificação, criada por Jamie Luguri e Lior Strahilevitz (2021) e reproduzida em tradução livre, que pode ser adotada como referência provisória para vislumbrar a diversidade dos *Dark Patterns*[25]:

Tabela 1 - Tipos de *Dark Patterns*

| Categoria | Variante | Descrição | Fonte |
|---|---|---|---|
| Provocativo | - | Pedidos repetitivos para incitar a parte a fazer alguma coisa | Gray *et al.* (2018) |
| Prova social | Mensagens de atividade; | Mensagens falsas/enganosas de que outros estão comprando ou contribuindo; | Mathur *et al.* (2019); |
| | Testemunhais. | Afirmações positivas falsas/enganosas de consumidores. | Mathur *et al.* (2019). |
| Obstrução | Hotel de baratas; | Assimetria entres os processos de inscrição e cancelamento; | Gray *et al.* (2018), Mathur *et al.* (2019); |
| | Prevenção de comparação | Frustra a comparação de | Brignull (2020), |

---

[25] Não se está alheio à variedade de taxonomias propostas por outros autores e à existência de *Dark Patterns* não contemplados pelo modelo de Luguri e Strahilevitz. No entanto, a tabela desenvolvida pelos pesquisadores é notória por sua atualidade e abrangência, além de conter notas descritivas sobre cada tipo de *Dark Pattern*, motivos pelos quais constitui um bom referencial ilustrativo do objeto de estudo. Com efeito, o objetivo da pesquisa é abordar os *Dark Patterns* enquanto objeto singular, abstraído de suas variadas espécies, de forma que a utilidade da tabela consiste na melhor visualização da temática e a compreensão de alguns tipos de Padrão recorrentemente referenciados pela doutrina, não se buscando um modelo taxonômico de alta precisão.

| | de preços; | preços durante a compra; | Gray *et al.* (2018), Mathur *et al.* (2019); |
|---|---|---|---|
| | Moeda intermediária | Compras realizadas em moedas virtuais para ocultar o preço real; | Brignull (2020); |
| | Contas imortais | As Conta e informações do consumidor não são passíveis de exclusão. | Bösch *et al.* (2016) |
| Sorrateiro | Esgueirar-se ao carrinho (*sneak into basket*); | O site adiciona itens não escolhidos pelo consumidor ao carrinho de compras; | Brignull (2020), Gray *et al.* (2018), Mathur *et al.* (2019); |
| | Custos ocultos; | Os custos são obscurecidos/escondidos até o final da transação; | Brignull (2020), Gray *et al.* (2018), Mathur *et al.* (2019); |
| | Inscrição oculta/ continuidade forçada; | Renovação automática não antecipada/indesejada; | Brignull (2020), Gray *et al.* (2018), Mathur *et al.* (2019); |
| | Isca-e-troca (*bait-and-switch*). | Venda de algo diferente do originalmente ofertado ao consumidor. | Gray *et al.* (2018). |
| Interferência de interface | Informação oculta | Informações importantes são visualmente ocultadas | Gray *et al.* (2018) |
| | Manipulação estética de pré-seleção; | Alternativas favoráveis ao fornecedor vêm pré-selecionadas | Bösch *et al.* (2016), Gray *et al.* (2018); |
| | Brincar com emoções; | Enquadramento emocionalmente manipulativo; | Gray *et al.* (2018); |
| | Falsa hierarquia/venda sob pressão; | Manipulação para incitar a escolha da versão mais cara; | Gray *et al.* (2018), Mathur *et al.* (2019); |
| | Pergunta capciosa; | Ambiguidade intencional ou óbvia; | Gray *et al.* (2018), Mathur *et al.* (2019); |
| | Anúncio disfarçado; | O consumidor é induzido a clicar em um anúncio que não se parece com um; | Brignull (2020), Gray *et al.* (2018); |
| | Culpabilizar (*confirmshaming*); | A escolha é enquadrada de uma forma que aparenta ser desonrosa ou estúpida; | Brignull (2020), Mathur *et al.* (2019); |
| | Fofura. | Robôs elaborados de forma atrativa para chamar consumidores. | Cherie & Catherine (2019). |
| Ação forçada | Spam amigo/pirâmide social/*address book leeching* | Extração manipulativa de informações sobre outros usuários; | Brignull (2020), Bösch *et al.* (2016), Gray *et al.* (2018); |
| | *Privacy Zuckering*; | Engana o consumidor a passar informações pessoais; | Brignull (2020), Bösch *et al.* (2016), Gray *et al.* (2018); |
| | Gamificação; | Funções desbloqueadas com o uso repetitivo; | Gray *et al.* (2018); |
| | Registro forçado. | Engana os consumidores a acharem que a inscrição é necessária. | Bösch *et al.* (2016). |
| Escassez | Mensagem de estoque baixo; | O consumidor é informado de quantidades limitadas; | Mathur *et al.* (2019); |
| | Mensagem de alta | O consumidor é informado | Mathur *et al.* |

| | demanda | que outros estão comprando o estoque remanescente | (2019); |
| Urgência | Contador de tempo; | Informação visual flagrante de que a oportunidade acabará em breve | Mathur *et al.* (2019); |
| | Mensagem de tempo limitado. | "A oportunidade acaba em breve" | Mathur *et al.* (2019). |

Fonte: LUGURI; STRAHILEVITZ, 2021, p. 53; tradução nossa.

Algumas inferências sobre as pesquisas citadas acima podem ser desde logo realizadas. Primeiro, trabalhos iniciais sobre o tema se debruçaram sobre a taxonomia dos *Dark Patterns*, e o fizeram a partir de uma literatura ainda recente, imatura e indisfarçavelmente inspirada pela nomenclatura do trabalho pioneiro de Harry Brignull (2010), resultando, como apontado por Mathur *et al.* (2021), em classificações imprecisas. Desta forma, apesar de um consenso geral sobre a existência dos *Dark Patterns* como espécie autônoma, a doutrina todavia não logrou em alcançar uma definição uníssona. Isto será explorado no capítulo 4 desta tese.

Uma segunda observação diz respeito à pequena quantidade de estudos jurídicos sobre os *Dark Patterns*, que ganharam maior expressividade apenas ao início da última década. É possível citar alguns trabalhos mais antigos, mas que não logram abordar a celeuma suficientemente. Por exemplo, um estudo introdutório sobre a taxonomia de "interfaces maliciosas" (*Malicious Interfaces*)[26], a tolerância à frustração dos usuários e as contramedidas usadas para delas se esquivar foi conduzido por Gregory Conti e Edward Sobiesk em 2010. O estudo de fato verificou um aumento na frustração desses indivíduos e que os usuários usam uma variada gama de contramedidas. Estes resultados, no entanto, são minados pelos baixos número e variedade de participantes[27] (CONTI; SOBIESK, 2010), concluindo os

---

[26] Ao momento da elaboração da pesquisa em comento, abril de 2010, o termo *Dark Patterns* todavia não se encontrava consolidado, visto que Harry Brignull o propôs alguns meses depois, em julho.

[27] A pesquisa acerca da tolerância de frustração de usuários foi conduzida com 27 participantes da conferência de *hackers* "HOPE" (*Hackers on Planet Earth*), 61 graduandos de uma faculdade americana e 47 participantes das conferências *BlackHat* e *Defcon*, de teor similar à da "HOPE" (CONTI; SOBIESK, 2010, p. 274-275), ou seja, um número de 135 participantes restritos a grupos sociais irrisoriamente homótonos, associados a um maior grau de escolarização e a atuações profissionais consideravelmente próximas ao objeto de estudo. De forma ainda mais restrita, as contramedidas utilizadas por usuários foram apuradas apenas com a entrevista aos 47 participantes das conferências *BlackHat* e *Defcon*,

próprios autores que mais estudos seriam precisos para entender as consequências do uso dessas interfaces (CONTI; SOBIESK, 2010, p. 278-279).

Além disso, a quase totalidade dos estudos jurídicos sobre as técnicas foi conduzida por instituições e pesquisadores americanos e de países integrantes da União Europeia, com atuações minoritárias de outros agentes, como o SERNAC chileno, sem a presença significativa de estudos brasileiros.

Por fim, uma terceira observação aponta que os efeitos decorrentes do uso de *Dark Patterns* são em grande parte desconhecidos. A preocupação que advém do uso dessas técnicas é baseada em bancos de amostra relativamente pouco abrangentes ou incompletos, como o rol de empresas analisado na pesquisa do SERNAC (2021) e os dados colhidos pelo *web crawler* criado por Mathur *et al.* (2019). O artigo escrito por Bösch *et al.* (2016) foi um dos poucos a trazer explicação psicológica técnica acerca do funcionamento dos *Dark Patterns*.

Os autores apontam a existência de dois tipos de sistemas cognitivos humanos que atuam nos processos de pensamento e tomada de decisão, os sistemas de tipo 1 e 2. O comportamento de Sistema 1 - despreocupado e automatizado - é encorajado quando os usuários "(a) possuem pouca motivação para pensar e arrazoar de maneira esforçada ou (b) não têm oportunidade para fazê-lo por causa da falta de conhecimento, habilidade ou tempo", apresentando-se como exemplo clássico a leitura de termos de uso e políticas de privacidade. Nestes casos, a "linguagem utilizada nos termos [...] são complicadas e os sujeitos são incapazes de interpretar essas informações", desencorajando pensamentos de tipo 2 - racionais e lentamente calculados (BÖSCH *et al.*, 2016, p. 246; tradução nossa)[28].

Valendo-se de conhecimentos sobre a cognição humana, os designers de UX/UI elaborariam os *Dark Patterns* com o intuito de forçar uma situação de Sistema 1, ou seja, para desmotivar o usuário a tomar ações reflexivas contrárias aos interesses da empresa, através, por exemplo, do uso de um timer para a finalização da compra ou de um anúncio sobre o final de uma promoção relâmpago. Da mesma forma, o emprego de mecanismos de inscrição que dependem de formas dificultosas de cancelamento desincentiva o consumidor a encerrar o vínculo, serviços que podem não mais ser de seu interesse.

---

[28] Do original: "Humans engage in System 1 processing whenever they (a) have little motivation to think and reason in an effortful way or (b) have no opportunity to do so because they lack the required knowledge, ability, or time. Users, for instance, often have no motivation to read general terms and conditions. In instances where they are motivated, they often do not have the opportunity to use System 2 thinking because the language used in general terms and conditions often is too complicated and subjects are unable to interpret this information".

Contudo, a ausência de estudos jurídicos e a falta de evidência da eficácia dos *Dark Patterns* foram desafiadas no que pode ser definido como uma segunda onda empirista de estudos, com forte representação nos trabalhos de Luguri e Strahilevitz (2021). Os autores conduziram experimentos para determinar precisamente quais são os impactos desses padrões na aquisição de produtos ou serviços, propondo

> um cenário *bait-and-switch* que pudesse ser visto pelos consumidores como plausível. Nos valeríamos de uma empresa de pesquisas para recrutar uma parcela da população americana adulta para participar de estudos de avaliação das atitudes adotadas pelos participantes acerca de privacidade. Após, enganaríamos esses adultos a acreditarem, ao final da pesquisa, que porque eles expressaram um forte interesse em privacidade (como os participantes tipicamente fazem em pesquisas) foram inscritos em um caro serviço de proteção contra roubos de identidade, fornecendo a opção de cancelar a inscrição. O experimento seria estruturado de forma a fazê-los acreditar que seu dinheiro estava em jogo, pois teriam que pagar pelo serviço se não optassem por encerrar a assinatura. Então eles variariam aleatoriamente se a oportunidade de desinscrever-se seria impeça por diferentes dosagens de *Dark Patterns*. (LUGURI, STRAHILEVITZ, 2021, p. 59; tradução nossa)[29]

Após uma seleção prévia e a coleta de dados demográficos, foram contabilizados 1.963 participantes imparciais. Para determinar se o preço constituiria um fator considerável para a eficácia dos *Dark Patterns*, foram utilizados dois valores mensais distintos na oferta do serviço, US\$2.99 ou US\$8.99. (LUGURI, STRAHILEVITZ, 2021, p. 59-62)

Informados da inscrição automática no serviço, os participantes receberam a opção de cancelá-la, mas em condições "suaves" ou "agressivas" de uso de *Dark Patterns* - determinadas conforme a quantidade de Padrões usados e a gravidade de suas restrições -, fora o grupo controle, no qual não se utilizou *Dark Patterns*. (LUGURI, STRAHILEVITZ, 2021, p. 59-62)

---

[29] "We designed a bait-and-switch scenario that would strike consumers as plausible. We would use an existing survey research firm to recruit large populations of American adults to participate in research studies that would evaluate their attitudes about privacy. Then we would deceive those adults into believing, at the end of the survey, that because they expressed a strong interest in privacy (as respondents typically do in surveys), we had signed them up for a costly identity theft protection service and would give them the opportunity to opt-out. We would structure the experiment in a way so as to make experimental subjects believe that their own money was at stake and they would need to pay for the service if they did not opt-out. Then we would randomly vary whether the opportunity to opt-out was impeded by different dosages or types of dark patterns"

Como resultado, somente 11,3% dos participantes no grupo controle aceitaram permanecer com o serviço de proteção. No grupo submetido a *Dark Patterns* suaves a aceitação subiu 228%, alcançando 25,8% destes participantes, enquanto a aceitação do serviço pelo grupo submetido a *Dark Patterns* agressivos quase quadruplicou, totalizando 41,9% destes participantes. A variação nos preços do serviço não interferiu significativamente nos índices de aceitação de qualquer um dos 3 grupos. (LUGURI; STRAHILEVITZ, 2021, p. 64-67)

Ao final do teste, os pesquisadores avaliaram o estado emocional dos participantes. Os dados não mostraram diferenças significativas entre o grupo controle e o grupo submetido a *Dark Patterns* em condição suave, mas sinalizaram um humor substancialmente negativo nos indivíduos expostos a *Dark Patterns* em condição agressiva. Esta diferença ocorreu especialmente nas hipóteses em que as pessoas recusaram o serviço, não se averiguando diferenças significativas de humor quando elas o aceitaram.

Os pesquisadores também mediram os impactos psicológicos dos Padrões através da quantidade de pessoas que deixaram a entrevista nas condições de grupo controle e de *Dark Patterns* suaves ou agressivos. Descobriu-se que os participantes sujeitos à modalidade agressiva foram muito mais propensos a deixar o experimento durante a fase de apresentação do *Dark Pattern* que nos demais – na condição agressiva, 65 deixaram o teste; no grupo de *Dark Patterns* suaves, apenas 9; e nenhum participante do grupo controle saiu.[30] (LUGURI; STRAHILEVITZ, 2021, p. 69-70)

Além disso, usuários submetidos a *Dark Patterns* também foram mais propensos a afirmar que sua liberdade no aceite ou recusa do serviço foi restringida. Não obstante, a maioria dos participantes em todos os grupos se sentiu mais livre que coagida. (LUGURI; STRAHILEVITZ, 2021, p. 69-70)

"Esses resultados sugerem que as empresas podem sofrer *backlash* e a perda da boa vontade dos consumidores se forem longe demais e apresentarem aos

---

[30] "Constatamos que os entrevistados eram muito mais propensos a desistir e se desvincular do estudo na condição agressiva [...]. Apenas nove participantes desistiram na condição leve, enquanto sessenta e cinco desistiram em algum momento na condição agressiva. Esta última é uma taxa de desistência incomum e surpreendentemente alta em nossa experiência, ainda mais significativa considerando a falácia dos custos irrecuperáveis. Os entrevistados normalmente dedicaram dez minutos ou mais à pesquisa antes de encontrar o *Dark Pattern* e, ao sair durante a parte do *Dark Pattern*, perderam dinheiro que poderiam muito bem ter sentido que já mereciam." (LUGURI, STRAHILEVITZ, 2021, p. 69)

seus clientes uma série gritante de *Dark Patterns*" (LUGURI; STRAHILEVITZ, 2021, p. 67; tradução nossa)[31], bem como o uso de Padrões de forma atenuada quase elimina o risco de efeitos negativos à imagem da empresa. (LUGURI, STRAHILEVITZ, 2021, p. 67-70)

A partir do recolhimento dos dados demográficos, identificou-se que indivíduos com níveis educacionais inferiores são mais propensos a aceitar os *Dark Patterns*. A educação não foi um elemento determinante para a aceitação dos *Dark Patterns* no grupo controle (LUGURI; STRAHILEVITZ, 2021, p. 70-71).

Um segundo experimento visou testar o desempenho de diferentes tipos de *Dark Patterns*. Os pesquisadores utilizaram uma narrativa similar à do primeiro experimento, mas desta vez empregando 5 condições de uso de *Dark Patterns* aplicadas à oferta[32] e outras 4 à maneira como os participantes aceitariam ou declinariam o serviço[33]. Em seguida, metade de todos os participantes foi submetida à pergunta capciosa "você prefere não declinar este serviço gratuito de proteção de dados e de monitoramento de crédito?" (tradução nossa)[34], à qual é possível responder "sim" ou "não" e cuja resposta para declinar o serviço é, contrariamente ao senso comum, "não" (LUGURI; STRAHILEVITZ, 2021, p. 73-74). Novamente houve variação dos valores mensais do serviço fictício para averiguar a sua influência nos efeitos dos *Dark Patterns*, mas desta vez os valores mensais foram fixados em US$8,99 e US$38,99. (LUGURI; STRAHILEVITZ, 2021, p. 76).

Os resultados foram compilados nas tabelas abaixo.

Tabela 2 - Aceitação na condição de conteúdo

| Condição | aceitação (%) | Número de aceites |
|---|---|---|
| Grupo controle | 14.8 | 191 (de 1.289) |
| Escassez | 14.3 | 91 (de 635) |
| Culpabilização/*confirmshammi* | 19.6 | 120 (de 612) |

---

[31] "These results suggest that if companies go too far and present customers with a slew of blatant dark patterns designed to nudge them, they might experience backlash and the loss of goodwill."
[32] Condições de grupo controle (sem *Dark Patterns*) e das espécies "escassez", "culpabilização" (*confirmshamming*), "prova social" e "informação oculta".
[33] Condições de grupo controle (sem *Dark Patterns*), "recomendação", "aceite por padrão" e "obstrução".
[34] "Would you prefer not to decline this free data protection and credit history monitoring?"

*ng*

| Prova Social | 22.1 | 140 (de 634) |
| Informação oculta | 30.1 | 183 (de 607) |

Fonte: LUGURI; STRAHILEVITZ, 2021, p. 76; tradução nossa

Tabela 3 - Aceitação na condição de aceite

| Condição | aceitação (%) | Número de aceites |
|---|---|---|
| Grupo controle | 16.7 | 216 (de 1.294) |
| Recomendação | 18.1 | 156 (de 861) |
| aceite por padrão | 20.1 | 171 (de 851) |
| Obstrução | 23.6 | 182 (de 771) |

Fonte: LUGURI; STRAHILEVITZ, 2021, p. 76; tradução nossa

Da análise das tabelas, revela-se que as porcentagens de aceitação do serviço nas condições de *Dark Patterns* foram superiores às dos grupos controle, apesar dos diferentes níveis de eficácia, com a exceção do grupo submetido aos Padrões de escassez. Isso mostra que as espécies de *Dark Patterns* influenciam os usuários em graus distintos, com a possibilidade de que alguns tipos não exerçam efeitos significativos, apesar da maioria o fazerem. Novamente o acréscimo dos preços não se mostrou um fator relevante, mesmo que um deles fosse desproporcionalmente maior que o outro. (LUGURI; STRAHILEVITZ, 2021, p. 76).

Questionados, 73,7% dos participantes que aceitaram o serviço avaliaram em 4 ou mais (de 7) a probabilidade de seu posterior cancelamento, e 21,1% destes assinalaram um 7. Os resultados deste experimento reiteraram a influência do nível educacional dos participantes na suscetibilidade aos *Dark Patterns*, observando-se maiores índices de aceitação por pessoas de baixa escolaridade.[35] (LUGURI; STRAHILEVITZ, 2021, p. 76-81; tradução nossa)

---

[35] "In short, although the results of Study 1 and 2 were not identical, both studies show that less education increases vulnerability to small- or moderate-dose dark patterns."

Outras pesquisas alcançaram resultados similares. O artigo de Blake *et al.* (2018), um dos maiores estudos já conduzidos na matéria[36], procurou compreender quais seriam os efeitos da ocultação do preço real de um bem até o final do processo de aquisição - o que caracteriza o emprego de um *Dark Pattern* de "custos ocultos" - em comparação com a transparência sobre os valores desde o início. Como raro diferencial em relação a outros trabalhos sobre *Dark Patterns*, a pesquisa foi realizada em um website autêntico (stubhub.com, uma plataforma de revenda de bilhetes). (BLAKE *et al*, 2018).

A pesquisa apurou que informar o valor integral do produto apenas ao final do processo de compra acarretou aumento de 8% nos lucros, quando o valor extra revelado somava 7,35% do total, e de 21%, quando o valor adicional somava 15% (BLAKE *et al*, 2018, p. 14). Assim, o aumento total nos lucros da empresa com a implementação do *Dark Pattern* ultrapassou 20% em comparação com optar pela transparência. (BLAKE *et al*, 2018, p. 32).

O artigo publicado por Sin *et al* (2022) apresentou mais demonstrativos experimentais de que alguns tipos de *Dark Patterns* ("quantidade limitada", "testemunhos falsos de consumidores" e "alta demanda") são capazes de fomentar comportamentos impulsivos de consumo, na medida que os graus de eficácia apurados são significativamente maiores que os do grupo controle não exposto a *Dark Patterns* (SIN *et al*, 2022, p. 8).

O estudo também avaliou a eficácia de métodos interventivos para combater esses *Dark Patterns* (SIN *et al.*, 2022). Foram testadas 3 técnicas: "adiamento" (*postponement*), no qual o consumidor é oferecido um intervalo para pensar sobre sua escolha; "reflexão" (*reflection*), que consiste em perguntar ao consumidor se a sua justificativa para tomar uma ação é válida; e "distração" (*distraction*), apresentando ao consumidor tarefas adicionais não relacionadas com a aquisição do produto, vez que, a *contrario sensu* do ordinariamente esperado, este método pode apresentar resultados benéficos aos consumidores se corretamente utilizado. (SIN *et al.*, 2022) Os testes revelaram influências variáveis, mas positivas, dos métodos interventivos para refrear os Dark Patterns estudados. (SIN *et al.*, 2022)

---

[36] De acordo com Harry Brignull, em manifestação na conferência "Trazendo Dark Patterns à Luz: um *Workshop* FTC" (*Bringing Dark Patterns to Light: An FTC Workshop*), de 2021.

Em arremate, os estudos apreciados neste capítulo revelam que *Dark Patterns* são objetos complexos de patente interesse jurídico, porquanto capazes de interferir gravemente na esfera jurídica dos usuários da interface, apesar de não existir consenso teórico sobre os contornos e efeitos dessas técnicas. No tocante à "primeira onda", focada em estudos primordialmente taxonômicos e classificatórios, ressaltou-se, com amparo na pesquisa de Mathur *et al* (2021), a discordância sobre os elementos integrantes de seu conceito, além de uma miríade de tipos de *Dark Patterns*. Uma "segunda onda" de pesquisas empíricas consistentemente demonstrou que *Dark Patterns* exercem efeitos significativos sob o comportamento dos usuários, em maior ou menor medida conforme os tipos e a quantidade de Padrões usados, e podem gerar benefícios aos seus implementadores. Além disso, a pesquisa de Sin *et al.* (2022) merece destaque ao trazer à tona a eficácia de métodos interventivos para combater os *Dark Patterns*.

## 3. PERSPECTIVAS SOBRE OS *DARK PATTERNS* EM OUTROS ORDENAMENTOS JURÍDICOS

Neste capítulo serão abordados posicionamentos e casos relevantes dos Estados Unidos da América, União Europeia e da OCDE frente aos *Dark Patterns*, considerando: a definição adotada para os caracterizar; a postura adotada em relação a sua juridicidade; e quais foram os fundamentos para disciplinar, ou manifestar-se de determinada maneira sobre os *Dark Patterns*.[37]

### 3.1 ANÁLISE DOS *DARK PATTERNS* SEGUNDO O DIREITO AMERICANO

Quiçá o país onde os debates sobre a juridicidade dos *Dark Patterns* tomaram força mais rapidamente é os EUA, cuja postura institucional é balizada principalmente pelos trabalhos da Comissão Federal de Comércio (*Federal Trade Comission*; tradução nossa), ou FTC, que despendeu vários esforços para a difusão de conhecimentos sobre os Padrões, sua regulamentação e fiscalização.

A disciplina dos *Dark Patterns* pela FTC teve como uma de suas primeiras expressões as atividades fiscalizatórias contra a empresa Age of Learning, Inc., que culminaram em demanda iniciada em 2020 junto à Corte dos Estado Unidos da América para o Distrito Central da Califórnia (tradução nossa)[38],[39]. Na ocasião, alegava-se que a plataforma de aprendizagem infantil online "ABCmouse", comercializada pela Age of Learning através de planos anuais, dificultava enormemente o cancelamento do serviço. Para efetuar a desinscrição, a plataforma online exigia que os consumidores navegassem por um sistema feito propositalmente complexo, repleto de telas com perguntas capciosas, anúncios disfarçados e insistentes tentativas de levar o consumidor a não cancelar o serviço (ESTADOS UNIDOS, 2020). Os planos anuais para a contratação do serviço também se renovavam automaticamente ao final do ano caso não houvesse o cancelamento (o que é chamado de *opt-out feature, negative option feature* ou, em tradução livre, funcionalidade ou sistema de opção negativa)[40], informação deixada de ser adequadamente repassada ao consumidor (ESTADOS UNIDOS, 2020).

---

[37] Se está ciente de que outros países já começaram a se debruçar sobre a matéria, como a Austrália, através de seu robusto relatório "Relatório investigativo provisório sobre plataformas digitais n.º 3: padrões de pesquisa e telas de escolha" (*Digital platform services inquiry Interim report No. 3 – Search defaults and choice screens*, em inglês). No entanto, instituições e órgãos dos EUA, Europa, União Europeia e OCDE foram escolhidos em razão de atualmente liderarem os debates sobre os *Dark Patterns* em âmbitos acadêmico e regulatório, como poucas organizações no mundo fizeram com tamanha qualidade, ou seja, a presente seção não visa contemplar todas as manifestações acerca dos *Dark Patterns*, e sim debruçar-se sobre uma monta limitada, mas quantitativa e qualitativamente significativa, de instâncias de posicionamento acerca da temática, a fim de definir um quadro geral sobre o tratamento conferido aos *Dark Patterns* pelo mundo.

[38] "United States District Court for the Central District of California"

[39] Segundo os registros contidos na ferramenta de pesquisa do website da *Federal Trade Comission*, este foi o primeiro caso noticiado amplamente pela FTC no qual a prática ilegal combatida foi identificada como um *dark pattern*.

Apresentadas as denúncias ao órgão, a FTC ingressou com demanda em face da empresa sustentando que a plataforma ABCmouse violaria a seção 5, (a), do "Ato da Comissão Federal de Comércio" (tradução nossa)[41], que veda a prática de atos enganosos[42], e o "Ato para Restaurar a Confiança dos Consumidores Online" (ROSCA) (tradução nossa)[43] em conjunto com a "Norma sobre Serviços de Telemarketing" (TSR) (tradução nossa)[44], que proíbem o uso de sistemas de opção negativa quando descumpridos os requisitos legais de transparência[45].

---

[40] De acordo com a "Regra sobre Serviços de Telemarketing" (tradução de *Telemarketing Services Rule*), §310.2, (w), *negative option feature*, ou funcionalidade de opção negativa "significa, em uma oferta ou acordo para o fornecimento de bens ou serviços, uma previsão na qual o silêncio do consumidor ou a falha em tomar uma atitude afirmativa para rejeitar bens ou serviços ou cancelar o acordo é interpretada pelo fornecedor como a aceitação da oferta." (Tradução nossa)

[41] "O *Federal Trade Comission Act* é o estatuto principal da Comissão. Nos termos desta Lei, a Comissão tem poderes, entre outras coisas, para (a) prevenir métodos desleais de concorrência e atos ou práticas desleais ou enganosas no, ou que afete o, comércio; (b) buscar reparação monetária e outras responsabilizações para condutas prejudiciais aos consumidores; (c) prescrever regras que definam com especificidade atos ou práticas desleais ou enganosas, e estabeleçam requisitos destinados a prevenir tais atos ou práticas; (d) recolher e compilar informações e conduzir investigações relativas à organização, negócios, práticas e gestão de entidades que se dedicam ao comércio; e (e) fazer relatórios e recomendações legislativas ao Congresso e ao público. Uma série de outros estatutos listados aqui são aplicados sob o *FTC Act*." (FTC, s.d.; tradução nossa)

[42] *FTC act*, §45, Seção 5, (a) e (1): "Declaração de ilegalidade; poder de proibir práticas desleais; inaplicabilidade ao comércio exterior. [...] Métodos desleais de concorrência no comércio ou que afetem o comércio, e atos ou práticas desleais ou enganosos que afetem o comércio, são declarados ilegais." (tradução nossa)

[43] O *Restore Online Shoppers' Confidence Act* (ROSCA) "proíbe que qualquer vendedor terceirizado pós-transação (um vendedor que comercializa bens ou serviços on-line por meio de um comerciante inicial depois que um consumidor iniciou uma transação com esse comerciante) cobre de qualquer conta financeira em uma transação na Internet, a menos que tenha divulgado claramente todo os termos materiais da transação e obteve o consentimento expresso e informado do consumidor para a cobrança. O vendedor deve obter o número da conta a ser cobrada diretamente do consumidor." (FTC, s.d.; tradução nossa)

[44] O *Telemarketing Sales Rule* (TSR) "exige que os operadores de telemarketing façam divulgações específicas sobre informações materiais; proíbe declarações falsas; estabelece limites nos horários em que os operadores de telemarketing podem ligar para os consumidores; proíbe chamadas para um consumidor que pediu para não ser chamado novamente; e estabelece restrições de pagamento para a venda de certos bens e serviços." (FTC, s.d.; tradução nossa)

[45] ROSCA, §8403: "É ilegal cobrar ou tentar cobrar de qualquer consumidor por quaisquer bens ou serviços vendidos em uma transação efetuada na Internet por meio de um recurso de opção negativa (conforme definido na Regra de Vendas de Telemarketing da Federal Trade Commission na parte 310

À luz destas acusações, a FTC e a Age of Learning firmaram acordo para encerrar a demanda, comprometendo-se esta com a correção das práticas abusivas - passando a recolher o consentimento expresso e informado dos usuários para o uso de funcionalidades de opção negativa e implementar mecanismos simplificados de cancelamento dos serviços. Adicionalmente, a empresa se obrigou a prestar contas sobre o cumprimento das determinações e pagar US$10.000.000,00 (dez milhões de dólares) em favor da FTC. (ESTADOS UNIDOS, 2020).

O termo "*Dark Pattern*" nenhuma vez foi mencionado para definir as técnicas usadas no programa ABCmouse durante o caso judicial, atendo-se estritamente aos conceitos de "atos enganosos" previstos em lei (ESTADOS UNIDOS, 2020). Sem embargos, a associação destas práticas com os *Dark Patterns* veio a ser feita no parecer de comissário da FTC, Rohit Chopra (2020), lavrado postumamente ao desfecho do caso.

No parecer, o comissário indicou que o Padrão empregado na ABCmouse foi um "hotel de baratas" (*roach motel*), serviço no qual um consumidor pode facilmente ingressar mas dificilmente sair, e que práticas como esta violam dispositivos da legislação esparsa sobre os direitos do consumidor, *e.g.*, os supramencionados ROSCA e Ato da FTC, além do "Ato para Controlar o Assalto de Pornografia e Marketing Não Solicitados" (CAN-SPAM) (tradução nossa)[46]. (CHOPRA, 2020)

A partir disso, verifica-se uma mobilização de esforços consideráveis da FTC para tratar dos *Dark Patterns*, firmando o seu posicionamento institucional através de eventos, com destaque à conferência intitulada "Trazendo Dark Patterns à Luz: um

---

do título 16 , Código de Regulamentos Federais), a menos que a pessoa— (1) forneça texto que divulgue de forma clara e visível todos os termos relevantes da transação antes de obter as informações de cobrança do consumidor; (2) obtém o consentimento expresso e informado do consumidor antes de cobrar do cartão de crédito, cartão de débito, conta bancária ou outra conta financeira do consumidor produtos ou serviços por meio de tal transação; e (3) fornece mecanismos simples para um consumidor impedir que cobranças recorrentes sejam feitas no cartão de crédito, cartão de débito, conta bancária ou outra conta financeira do consumidor." (tradução nossa)

[46] O *Controlling the Assault of Non-Solicited Pornography and Marketing Act* (CAN-SPAM) "estabelece requerimentos para aqueles que enviarem e-mails comerciais não solicitados. O Ato bane títulos falsos ou enganosos e proíbe o uso de mensagens capciosas. Ele também requer que e-mails comerciais não solicitados sejam identificados como anúncios e providenciem aos receptores um método para optar por não mais receber e-mails como este no futuro. Adicionalmente, o Ato direciona à FTC a competência para emitir regras sobre a rotulação de e-mails com conteúdo sexual explícito como tais e estabelecimento de critérios para a determinação dos propósitos primários de um e-mail comercial". (FTC, s.d.; tradução nossa)

*Workshop* FTC"[47] (2021; tradução nossa)[48], e ao relatório Combatendo Males Online através de Inovação" (2022; tradução nossa)[49]. Ambos consolidam opiniões contrárias aos *Dark Patterns*[50], identificando-os como meios coercitivos para obter dados pessoais e ofertar produtos e serviços (FTC, 2022). Esse entendimento pode ser observado nos subsequentes processos judiciais movidos pela Comissão, entre os quais se destaca o caso FTC v. Credit Karma, LLC.

No processo encerrado em setembro de 2022, a empresa Credit Karma, LLC. foi acusada de empregar *Dark Patterns* no sistema de aquisição de cartões de crédito. Após os consumidores finalizarem o cadastro no sistema compartilhando seus dados pessoais, automaticamente disparavam-se mensagens informando que haviam sido, ou possuiriam enormes chances de ser, pré-aprovados para receber o produto (ESTADOS UNIDOS, 2022). No entanto, diminutos asteriscos e notas de rodapé nessas mensagens remetiam à informação de que a Credit Karma não pré-aprovava seus clientes, a despeito do fortemente insinuado pelas frases em destaque.

Isto resultou no indeferimento de cerca de um terço dos consumidores inscritos para obter o cartão, ferindo sua legítima expectativa e gastando seu tempo útil. Além disso, os dados pessoais dos consumidores foram reapropriados para satisfazer finalidades diversas da informada, incluindo o envio de anúncios por e-mail (ESTADOS UNIDOS, 2022).

Assim, configurado trespasse à seção 5 do ato da FTC, as partes alçaram acordo para, entre outras obrigações, condenar a empresa Credit Karma a encerrar as práticas ilegais e pagar US$3.000.000,00. (ESTADOS UNIDOS, 2022)

Ainda no âmbito da regulação federal dos *Dark Patterns*, o Projeto de Lei do "Ato para a Redução de Experiências Enganosas a Usuários Online" (DETOUR) (tradução nossa)[51] será um importante avanço na disciplina dessas técnicas, caso aprovado pelo congresso americano.

---

[47] Os debates conduzidos neste evento foram sintetizados no relatório "Trazendo *Dark Patterns* à Luz" (*Bringing Dark Patterns to Light*, em inglês) publicado no mês de setembro de 2022, que solidifica o posicionamento da FTC de forma contrária às práticas (FTC, 2022).
[48] "Bringing Dark Patterns to Light: An FTC Workshop"
[49] "Combating Online Harms Through Innovation"
[50] Este posicionamento se fez marcante no "*Enforcement Policy Statement Regarding Negative Option Marketing*" ("Declaração de Política Sobre a Regulação do Marketing de Opção Negativa", em tradução livre), o qual reitera a posição legal defendida no processos *FTC v. Age of Learning, Inc.,* de que ocorreria violação ao *FTC act, ROSCA e ao TSR*
[51] "Deceptive Experiences To Online Users Reduction Act (DETOUR Act)"

Não há menção literal aos *Dark Patterns* no DETOUR, optando o legislador, alternativamente, por se referir a "Interface de usuário com o propósito ou substantivo efeito de obscurecer, subverter, ou debilitar a autonomia, capacidade decisória, ou escolha dos usuários para obter consentimento ou dados" (DETOUR, Sec. 3, (a), (1); tradução nossa)[52][53], que for implementada por Grandes Operadores Online (*Large Online Operators*, ou LOOs) - empresas sujeitas à jurisdição da FTC e que fornecem serviços online para mais de 100 milhões de clientes a cada 30 dias.

Em síntese, o DETOUR impõe às LOOs a obrigação de não manipular interfaces de usuários das maneiras especificadas na norma (vide nota n.º 53) e de prestar esclarecimentos, a cada 90 dias, sobre experimentos psicológicos ou comportamentais realizados em suas plataformas online. Os experimentos deverão ser descontinuados caso averigue-se que o consentimento obtido como resultado é insuficiente ou não poderá ser recolhido no prazo de 2 dias úteis. As LOOs também precisarão manter conselho independente para analisar a obediência destas pesquisas à legislação.

Ante o descumprimento das prescrições trazidas acima, o DETOUR estabelece que a conduta representará malferimento às disposições do "ato da FTC" sobre práticas injustas ou deceptivas, conforme a seção 18 desta norma (15 U.S.C. 57a(a)(1)(B)), legitimando a *Federal Trade Comission* para apreciar o caso.

Algumas facetas do DETOUR foram incorporadas ao projeto do "Ato sobre Dados Seguros" (tradução nossa)[54] (FAZLIOGLU, 2020). Tal como no DETOUR, não há menção expressa ao termo "*Dark Pattern*", apenas se referenciando que atos e práticas injustos ou enganosos ferem a seção 18 do Ato da FTC, com a respectiva aplicação das sanções cabíveis.

---

[52] "Unfair and Deceptive Acts and Practices Relating to the Manipulation of User Interfaces".

[53] Três definições integram o conceito e aprofundam o rol de condutas proibidas (*prohibited conducts*) relacionadas aos *Dark Patterns*. Nos termos da Sec. 3, (a), (1), do DETOUR:"É ilegal que qualquer grande operador online [...] [1] projete, modifique ou manipule uma interface de usuário com propósito ou apresentando substanciais efeitos de obscurecer, subverter ou prejudicar a autonomia, capacidade decisório ou a escolha para obter consentimento para o tratamento de dados de usuários; [2] subdivida ou segmente consumidores de serviços online em grupos com o propósito de conduzir experimentos ou pesquisas comportamentais ou psicológicas, exceto com o consentimento informado de cada usuário envolvido; [3] ou projete, modifique ou maipule uma interface de usuário de um website ou serviço online, ou porção do mesmo, que seja direcionada a criança, com o propósito de causar, aumentar ou encorajar o uso compulsivo, inclusive de funções de repetição automática de vídeos iniciada sem o consentimento de um usuário." (tradução nossa)

[54] "Safe Data Act".

Além das medidas em nível federal, os Estados dos EUA também participam significativamente da regulamentação dos *Dark Patterns*, com destaque à Califórnia por seu "Ato sobre a Proteção da Privacidade dos Consumidores da Califórnia de 2018" (CCPA) (tradução nossa)[55], emendado pelo "Ato sobre os Direitos à Privacidade dos Consumidores da Califórnia" (CPRA) (tradução nossa)[56].

De largada, é necessário advertir que o uso de *Dark Patterns* parece violar, *ipso facto*, alguns dispositivos gerais encontrados nas leis de proteção de dados pessoais. Com efeito, a aplicação dos Padrões na coleta de dados afeta uma série de direitos tipicamente previstos em legislações do modelo europeu de proteção de dados pessoais[57], o qual largamente inspira o CCPA e o CPRA (PARDAU, 2018). A utilização de *Dark Patterns* configura afronta aos direitos de acesso (Cal. Civ. Code, §1798.100), deletar informações (Cal. Civ. Code, §1798.105), saber quais dados estão sendo tratados (Cal. Civ. Code, §1798.110 e ss.), e, em proporção ainda mais expressiva, ao direito de *opt-out* - de optar por cancelar a prestação de um serviço ou tratamento de dados pessoais (Cal. Civ. Code, §1798.120).

Possibilitando o enquadramento jurídico dos *Dark Patterns* perante a proteção de dados, o CCPA trabalha de forma genérica, ao passo que o CPRA os ataca frontalmente, sendo a primeira legislação a mencioná-los expressamente em seu bojo. O CPRA os define como "interface de usuário desenhada e manipulada com o substancial efeito de subverter ou limitar a autonomia, liberdade de decidir ou de escolher do usuário, conforme definido em lei." (Cal. Civ. Code, §1798.140, l; tradução livre)[58]. Combatendo-os, dispõe que não constitui consentimento o aceite obtido através de *Dark Patterns* (Cal. Civ. Code, §1798.140, h), ao mesmo tempo que veda o seu uso em páginas web que permitam o ingresso em serviços ou tratamentos de dados (*opt-in*) (Cal. Civ. Code, §1798.185, a, 20, C, ii).

---

[55] "California Consumer Privacy Act of 2018".

[56] "California Privacy Rights Act".

[57] "O modelo europeu, sistemático, estruturou-se primeiramente em torno de uma Diretiva, uma disciplina ampla e detalhada a ser transposta para a legislação interna de cada estado membro, e hoje está ordenado basicamente pelo Regulamento Geral de Proteção de Dados (GDPR, a sigla em inglês pela qual é internacionalmente reconhecido). O modelo norte-americano, por outro lado, apresenta-se fracionado, com disposições legislativas e jurisprudenciais concorrentes em uma complexa estrutura federativa, o que torna sua leitura em chave sistemática – e até mesmo a compreensão geral de seu conjunto – um desafio para os próprios juristas norte-americanos." (DONEDA, 2021, p. 191-192)

[58] ""Dark pattern" means a user interface designed or manipulated with the substantial effect of subverting or impairing user autonomy, decisionmaking, or choice, as further defined by regulation."

Outra importante legislação da Califórnia relevante à temática foi aprovada em 15 de setembro de 2022, o "Ato do Código de Design Apropriado para Menores da Califórnia" (AADC) (tradução nossa)[59]. O AADC veda o "uso de dark patterns para levar ou encorajar crianças a fornecer dados pessoais além do que é razoavelmente esperado para oferecer o serviço online, produto ou funcionalidade com o intuito de renunciar suas proteções à privacidade ou tomar qualquer ação que a empresa saiba, ou deveria saber, ser materialmente prejudicial à sua saúde física, saúde mental ou bem-estar" (Cal. Civ. Code, §1798.99.31, a, 7; tradução nossa)[60].

Sem embargos à proeminência legislativa da Califórnia sobre os *Dark Patterns*, outros Estados estão se mobilizando para dispor sobre assunto, alguns deles já possuindo norma específica a respeito. O Colorado, por exemplo, por meio do "Ato sobre Privacidade do Colorado" (CPA) (tradução nossa)[61], traça nomenclatura virtualmente idêntica à do CPRA (CPA, 6-1-1303, (9)), vedando a coleta do consentimento com o uso de *Dark Patterns* (CPA, 6-1-1303, (5), (c)).

Já no Estado da Virgínia o "Ato sobre a Proteção de Dados dos Consumidores da Virgínia" (VCDPA) não regula os *Dark Patterns*, mas aplicam-se os mesmos comentários feitos ao CCPA californiano: na medida que o VCDPA traça regras de proteção de dados, as técnicas sob estudo não possuiriam guarida diante dessa norma.[62]

## 3.2 ANÁLISE DOS *DARK PATTERNS* SEGUNDO O DIREITO DA UNIÃO EUROPEIA.

No direito europeu, os *Dark Patterns* são tópico de debate bastante recente, de forma que não surprenderia se a sua abordagem jurídica fosse incipiente. Contudo, muito pelo contrário, as discussões circundando a temática são de alto calibre, sendo possível verificar atividade legiferante em prol da regulação dos *Dark*

---

[59] "California Age-Appropriate Design Code Act"
[60] "Use dark patterns to lead or encourage children to provide personal information beyond what is reasonably expected to provide that online service, product, or feature to forego privacy protections, or to take any action that the business knows, or has reason to know, is materially detrimental to the child's physical health, mental health, or well-being."
[61] "California Privacy Act".
[62] "Virginia Consumer Data Protection Act".

*Patterns*, apesar de ainda restarem dúvidas sobre como conferir-lhes a abordagem apropriada.

Um grande passo para a tratativa dos *Dark Patterns* em solo europeu foi as "Diretrizes 03/2022 sobre *Dark Patterns* nas interfaces de plataformas de mídias sociais: como reconhecê-los e evitá-los" (2022) (tradução nossa)[63], do "Comitê de Proteção de Dados Europeu" (EDPB) (tradução nossa)[64]. É um importante documento que, a despeito de ausência de eficácia normativa, firma a visão do Comitê e entrelaça os estudos elaborados por autoridades de alguns países europeus, como França, Noruega e Reino Unido[65].

Estas Diretrizes conceituam os *Dark Patterns*, no contexto das redes sociais, como "interfaces e experiências de usuário implementadas em plataformas de mídia social que levam os usuários a tomar decisões não pretendidas, indesejadas e potencialmente danosas acerca do tratamento de seus dados pessoais" (EDPB, 2022, p. 2; tradução nossa)[66], estabelecendo, desde logo, a impossibilidade de adequar estas técnicas com o princípio do "processamento justo" (*fair processing*)[67].

---

[63] "Guidelines 3/2022 on Dark patterns in social media platform interfaces: How to recognise and avoid them".

[64] "European Data Protection Board"

[65] Estas nações contribuíram para os debates sobre os *Dark Patterns* através de suas autoridades de proteção de dados e direito do consumidor, com destaque à CNIL francesa (*Commission Nationale Informatique & Libertés*), à Forbrukerrådet Norueguesa, e à CMA Britânica (*Competition & Markets Authority*).

[66] "interfaces and user experiences implemented on social media platforms that lead users into making unintended, unwilling and potentially harmful decisions regarding the processing of their personal data".

[67] "[...] É um princípio abrangente que demanda os dados pessoais de titulares não sejam tratados de forma prejudicial, discriminatória, inesperada ou enganosa. Se uma interface fornece informações insuficientes ou enganosas para um usuário e preenche as características dos *Dark Patterns*, isto pode ser classificado como processamento injusto. O princípio da justiça possui uma função guarda-chuva e todos os *Dark Patterns* não podem com ele se adequar, independentemente da conformidade com outros princípios da proteção de dados pessoais." (EDPB, 2022, p. 8-9; tradução nossa) No original: "[...] Is an overarching principle which requires that personal data shall not be processed in a way that is detrimental, discriminatory, unexpected or misleading to the data subject. If the interface has insufficient or misleading information for the user and fulfils the characteristics of dark patterns, it can be classified as unfair processing. The fairness principle has an umbrella function and all dark patterns would not comply with it irrespectively of compliance with other data protection principles."

Adicionalmente, a EDPB também coloca os *Dark Patterns* em cheque perante "os princípios de prestação de contas[68], transparência[69] e a obrigação de proteção de dados por design [*privacy by design*][70] estabelecida no artigo 25 do RGPD"[71] (EDPB, 2022, p. 3; tradução nossa), recomendando que os controladores os levem em consideração quando averiguarem a existência de *Dark Patterns* em suas plataformas[72], sem prejuízo das variadas disposições legais específicas do RGPD (EDPB, 2022, p. 8-9).

As formas de consentimento são tidas com bastante significância pelo documento, enfatizando-se os momentos de inscrição do usuário na plataforma e de retirada do consentimento. O consentimento para a inscrição dos dados do usuário na plataforma precisa ser obtido de forma "livre, específica, informada e não-ambígua" (EDPB, 2022, p. 12; tradução nossa)[73], utilizando linguagem clara e acessível, consoante os requisitos do art. 7 do RGPD (EDPB, 2022, p. 12-14). O documento sublinha que a obtenção e a revogação do consentimento deverão ser simétricos, ou

---

[68] "Tido como a espinha dorsal das responsabilidades do controlador" segundo o RGPD (BESEMER, 2020, p. 34; tradução nossa), o princípio da prestação de contas se refere, de acordo com o RGPD, à obrigação de um controlador se responsabilizar e demonstrar a conformidade com os demais princípios da proteção de dados pessoais. Este se aplica às plataformas de mídia social em relação à demonstração de obediência às normas de proteção de dados, apresentando ao usuário de que forma os processos de tratamento impactam o uso de dados, por exemplo.

[69] O princípio da transparência é previsto no art. 5, (1), (a), do RGPD, mas não é acompanhado de uma definição. Sem embargos, o Considerando 39 do RGPD explana que: O princípio da transparência exige que as informações ou comunicações relacionadas com o tratamento desses dados pessoais sejam de fácil acesso e compreensão, e formuladas numa linguagem clara e simples. Esse princípio diz respeito, em particular, às informações fornecidas aos titulares dos dados sobre a identidade do responsável pelo tratamento dos mesmos e os fins a que o tratamento se destina, bem como às informações que se destinam a assegurar que seja efetuado com equidade e transparência para com as pessoas singulares em causa, bem como a salvaguardar o seu direito a obter a confirmação e a comunicação dos dados pessoais que lhes dizem respeito que estão a ser tratados.

[70] Por sua vez, o princípio da proteção de dados por design, mais conhecido por privacy by design, é extensamente caracterizado no art. 25 do RGPD, sendo possível sintetizá-lo como o mandamento que "obriga o controlador a implementar medidas técnicas e operacionais fim-a-fim completas" necessárias para " integrar as salvaguardas no tratamento para proteger os direitos dos titulares" (BESEMER, 2020, p. 36; tradução nossa). Sete são os elementos moldantes do Privacy by Design, segundo o magistério de Ann Cavoukian (2006; tradução nossa): "Proativo, não reativo; preventivo, não reparativo"; "Privacidade por Definição"; "Privacidade embutida no design"; "Funcionalidade plena - soma-positiva, não soma-zero"; "Segurança fim-a-fim - segurança do ciclo de vida"; "Visibilidade e transparência"; "Respeito pela privacidade do usuário".

[71] RGPD aqui se refere ao Regulamento Geral de Proteção de Dados, ou, em inglês, *General Data Protection Regulation* (GDPR), norma responsável por lançar as bases da proteção de dados pessoais na União Europeia.

[72] "[...] Accountability, transparency and the obligation of data protection by design stated in Article 25 GDPR "

[73] "freely given, specific, informed and [an] unambiguous indication of the data subject's wishes by which he or she, by statement or by a clear affirmative action, signifies agreement to the processing of personal data relating to him or her"

seja, é preciso ser possível fazê-la de maneira tão fácil como na inscrição (EDPB, 2022, p. 14).

Neste giro, de acordo com as Diretrizes 03/2022, Os *Dark Patterns* de "requerimentos insistentes", que integrariam o grupo de "*Dark Patterns* baseados em conteúdo" (EDPB, 2022, p. 14; tradução nossa)[74], afrontariam as normas sobre o consentimento ao instarem o usuário a tomar uma atitude após incessantes tentativas, ferindo, no exemplo dado pela EDPB (requerimento insistente do número de telefone para fins de segurança), os princípios da necessidade e finalidade. Outros *Dark Patterns* que impactam o momento da inscrição dos titulares incluem "informações enganosas" (*misleading information*), "direcionamento emocional" (*emotional steering*) e "desnecessariamente longo" (*longer than necessary*), extensamente definidos e condenados perante as regras e princípios da proteção de dados pessoais.

Outras considerações foram tecidas quanto à prestação de informações aos usuários, sua proteção, a concretização de direitos dos titulares de dados pessoais e o cancelamento de contas de usuário, em todas estas ressaltando-se os impactos negativos dos *Dark Patterns*. (EDPB, 2022)

Fora o posicionamento do EDPB exarado nas Diretrizes 03/2022, valiosos e ricos apontamentos sobre os *Dark Patterns* também são extraídos dos comentários a este documento - feitos por empresas, organizações e instituições diversas, entre as quais a empresa Meta, as Universidade de Luxemburgo e Chicago, organizações de pesquisadores sobre o mercado (EFAMRO e ESOMAR) e organizações de defesa do consumidor (BEUC e VZBV).

Discorrer sobre cada um dos 26 pareceres individualmente[75] foge ao escopo desta tese, mas é pertinente, e cabível, apanhar as principais considerações levantadas ao longo dos textos, com atenção especial aos pontos de vista de entidades que outrora não puderam participar com maior expressividade dos debates coletivos sobre os *Dark Patterns*.[76]

---

[74] "Content-based patterns".

[75] Fizeram comentários as seguintes entidades e indivíduos: *Spolek pro ochranu osobních údajů*, Universidade de Luxemburgo; *European Center for Digital Rights* (noyb); *Data & Marketing Association Finland*; *Interactive Advertising Bureau Europe* (IAB); Amurabi SAS; Meta Platforms Ireland Limited; BEUC; *Access Now, Simply Secure*; *World Wide Web Foundation*; Zsolt Bártfai; Luiza Jarovsky; EFAMRO; ESOMAR; *SME Connect - Coalition of Digital Ads of SMEs*; FEDMA; *Division Information and Consulting, Austrian Federal Economic Chamber*; Yuki Yuminaga; dacuro GmbH; Marco Costantini; Aleksandrs; Universidade de Chicago; Micha Beetz; estudantes do curso de mestrado em "lei e tecnologia na Europa" da Universidade de Utrecht; *Verbraucherzentrale Bundesverband* (vzbv); e dois indivíduos anônimos.

A leitura dos comentários revela que a percepção de empresas, entidades e indivíduos consultados é majoritariamente negativa, compreendendo-se de forma praticamente unânime que os *Dark Patterns* afetam transversalmente uma multiplicidade de direitos - como os do consumidor, de proteção de dados pessoais, o direito de empresa e da concorrência. No entanto, os termos nos quais a regulação deve ser erigida foram alvos de diversas críticas, demonstrando a ausência de unicidade teórica sobre essas técnicas. (UNIÃO EUROPEIA, 2022)

De início, observa-se que as opiniões sobre o próprio conceito jurídico de *Dark Patterns* são dissidentes, não havendo consenso sobre qual seria a definição apropriada, ao estilo do transposto sobre o trabalho de Mathur *et al.* (2021) no capítulo passado. Foram propostos critérios distintos para modificar o conceito dado pelas diretrizes ou, alternativamente, conferir-lhe redações totalmente distintas, (EUNIÃO EUROPEIA, 2022).

A partir dessa dissonância, os liames do que se consideraria um *Dark Pattern* em um contexto prático foram postos em dúvida, sobretudo no tocante aos exemplos dados pela EDPB. *Verbi gratia*, classificar como *Dark Patterns* a reiterada solicitação do número celular de titulares, para fins de proteção da conta de usuário, e o condicionamento do ingresso do usuário na rede social ao aceite dos termos de uso e política de privacidade gerou muitas discordâncias, encontrando-se justificativas válidas para a manutenção da legalidade destas práticas. Com efeito, no primeiro caso, a despeito da insistência, há uma preocupação legítima com a defesa da conta de usuário e, no segundo caso, há um vínculo contratual amparado pela legislação consumerista, existindo um natural desequilíbrio entre as partes que não é sozinho capaz de retirar a validade da obrigação. (UNIÃO EUROPEIA, 2022)

Muito exigiu-se o reconhecimento da vulnerabilidade de indivíduos perante os *Dark Patterns*, em particular das pessoas idosas, crianças e pessoas com menor grau de instrução, com o intuito de adequar a tutela jurídica a estas peculiaridades. Neste sentido, recomendou-se que a EDPB despenda esforços em conjunto com outras

---

[76] Para os fins desta análise, serão excluídos comentários referentes à estruturação e metodologia aplicados às diretrizes 03/2022 e a temas de proteção de dados pessoais fora do contexto dos *Dark Patterns*, vez que não guardam afinidade temática com o objeto sob estudo.

entidades especializadas na defesa do consumidor, e das referidas minorias, para atacar a celeuma sob ângulos distintos. (ALEKSANDRS; *et al*, 2022)

Por fim, foi levantada a preocupação de que o controle excessivo dos *Dark Patterns*, tendo em vista as dificuldades conceituais atreladas às técnicas, comprometeria a concorrência e o exercício do direito de empresa, levando as interfaces gráficas a um estado de uniformidade incompatível com o fomento à inovação. Também recomendou-se que a EDPB ilustre quais práticas seriam permitidas ou encorajadas - o que a Universidade de Chicago chamou de *Bright Patterns* -, com o intuito de efetivamente instruir os designers sobre quais caminhos trilhar. (UNIÃO EUROPEIA, 2022)

Além das Diretrizes e os seus comentários, a Europa possui outras experiências bastante inovadoras no tratamento dos *Dark Patterns,* com destaque à atividade legislativa. Fora as normas sobre proteção de dados e as alusões ao direito de defesa do consumidor e da concorrência, a União Europeia promulgou, em 2022, importantes marcos legais sobre serviços prestados na internet: o "Regulamento Serviços Digitais" (*Digital Services Act*, ou DSA) e o "Regulamento Mercados Digitais" (*Digital Markets Act*, ou DMA), que normatizaram dispositivos sobre o abuso de interfaces de usuário - nas quais se encaixam os *Dark Patterns*.

O DMA (Regulamento Mercados Digitais) busca disciplinar as atividades dos "controladores de acesso", grandes empresas inseridas na internet que se encaixam nos requisitos de seu art. 1º.[77]

Uma vez enquadrados neste escopo, os controladores de acesso deverão obedecer às variadas regras dos arts. 5º e 6º do DMA - entre as quais é possível sublinhar as vedações à combinação e ao uso de dados não públicos gerados pelas atividades de utilizadores profissionais, como comércios eletrônicos hospedados em perfis de redes sociais - sem que, no entanto, haja uma norma direta sobre os *Dark Patterns* nesse âmbito.

A regra pertinente em relação aos Padrões é a constante no art. 11 do DMA, a cláusula de "antievasão", a partir do que a obediência aos artigos 5º e 6º "não pode ser prejudicada por comportamentos da empresa à qual o controlador de acesso pertence, independentemente desse comportamento ser de natureza contratual, comercial, técnica ou qualquer outra"[78]. Os controladores de acesso são igualmente obriganos a respeitar os padrões de proteção de dados exigidos para a coleta do consentimento e a não "deteriorar as condições ou a qualidade de nenhum dos serviços essenciais de plataforma [...] nem dificultar indevidamente o exercício desses direitos ou escolhas".

---

[77] *In verbis*: Um prestador de serviços essenciais de plataforma deve ser designado como controlador de acesso se: a) Tiver um impacto significativo no mercado interno; b) Explorar um serviço essencial de plataforma que serve de porta de acesso importante para os utilizadores profissionais chegarem aos utilizadores finais; e c) Ocupar uma posição enraizada e duradoura nas suas operações ou se for previsível que venha a ocupar tal posição num futuro próximo.

conforme salientado pelo BEUC[79] em seu relatório acerca dos *Dark Patterns* (2022), trata-se de um dispositivo que regula a matéria de forma indireta, restringindo a aplicação destas técnicas através da proibição do uso de meios técnicos - *v.g.*, interfaces maliciosas - para dificultar o cumprimento do DMA e o exercício de direitos e escolhas dos usuários, ratificando, noutra senda, o dever de cumprir com os requisitos legais do consentimento, o que é obstado por alguns tipos de *Dark Patterns*.

Por outro lado, o DSA (Regulamento Serviços Digitais) emprega medidas diretas para combater os *Dark Patterns*, visto que, de acordo com o seu art. 25a, 1, "provedores de plataformas online não devem projetar, organizar ou operar suas interfaces online para enganar, manipular ou materialmente distorcer ou prejudicar a habilidade dos receptores do serviço de tomar decisões livres e informadas" (tradução nossa)[80]. Não foi utilizado o termo *Dark Pattern* no texto final da norma, mas o parlamento europeu (2022) o empregou para se referir ao objeto regulado pelo artigo supra. O termo também foi usado no recital 39a da lei em comento, sendo possível afirmar seguramente que se está diante de vedação expressa aos *Dark Patterns* na União Europeia.

## 3.3 O POSICIONAMENTO DA OCDE

O último posicionamento relevante que será abordado é o da Organização para a Cooperação e o Desenvolvimento Econômico (OCDE), que orquestrou uma mesa redonda para debater os *Dark Patterns* e seus impactos sob os consumidores a nível internacional, originando abrangente e completo relatório.

---

[78] DMA, art. 11, 1: Os controladores de acesso devem assegurar o cumprimento efetivo e integral das obrigações previstas nos artigos 5.º e 6.º. Embora as obrigações previstas nos artigos 5.º e 6.º sejam aplicáveis no respeitante aos serviços essenciais de plataforma designados nos termos do artigo 3.º, a sua aplicação não pode ser prejudicada por comportamentos da empresa à qual o controlador de acesso pertence, independentemente de esse comportamento ser de natureza contratual, comercial, técnica ou qualquer outra.

[79] "Organização Europeia de Consumidores", ou, em Inglês, "The European Consumer Organization".

[80] "Providers of online platforms shall not design, organize or operate their online interfaces in a way that deceives, manipulates or otherwise materially distorts or impairs the ability of recipients of their service to make free and informed decisions."

Iniciando com a definição dos *Dark Patterns*, a OCDE salientou a existência das anteriormente mencionadas dificuldades conceituais, apontando não existir um consenso sobre a temática. A despeito disso, foram acatados os argumentos de Mathur *et al.* (2021) sobre as formas pelas quais os *Dark Patterns* se diferenciam de outras técnicas de marketing: por (i) interferirem na "arquitetura de escolha" dos usuários através de uma gama de atributos de design, modificando ou manipulando os fluxos de informação para restringir o número ou dificultar a tomada de escolhas livres e informadas; (ii) e porque são capazes de causar prejuízos ao "bem-estar" (tradução nossa)[81] individual e coletivo de consumidores e à sua autonomia individual (OCDE, 2021, p. 5).

A Organização reconheceu que as proporções do uso dos *Dark Patterns* tomaram patamares avassaladores, conclusão retirada a partir de pesquisas que atestam o amplo emprego destas técnicas em websites. Estes estudos também verificaram a sua prejudicialidade aos consumidores, na medida que exploram vieses cognitivos humanos para guiá-los ao comportamento desejado pelo implementador do *Dark Pattern* (OCDE, 2021, p. 6).

Ao final, a Organização ressaltou que, apesar da maioria dos estudos empíricos sobre *Dark Patterns* terem sido conduzidos por pesquisadores individuais, as entidades de defesa do consumidor possuem um vasto arcabouço de recursos para apreciar estas técnicas (OCDE, 2021, p. 7), devendo ser encorajadas a fazê-lo. Neste sentido, enfatizou-se que, porquanto o tema ainda não foi suficientemente regulado, apesar de alguns países pertencentes à OCDE já estabelecerem vedações a algumas das práticas entendidas como *Dark Patterns*, não existem normas suficientes que tratam a temática de forma unitária e ciente de suas particularidades, mesmo considerando as recentes iniciativas de autoridades de defesa do consumidor e proteção de dados (OCDE, 2021, p. 7).

---

[81] "Bem-estar", traduzido de "*Welfare*", possui conotação jurídica no direito costumeiro muito distinta do significado literal em português ("estado de satisfação plena das exigências do corpo e/ou do espírito.", segundo a Oxford Languages). O termo *Welfare*, segundo Bent Greve (2008), é complexo e possui diversos significados conforme o contexto em que é empregado, podendo referir-se ao estado econômico ou social de indivíduos, em um sentido "micro", ou ao "Produto Interno Bruto e total de gastos com políticas públicas de bem-estar (como indicador de recursos)" (GREVE, 2008, p. 59; tradução nossa), em um sentido "macro", entre outros conceitos, utilizado sozinho ou em conjunto a outras palavras, relacionados a institutos particulares do direito americano ou anglo-saxão (GREVE, 2008). Sem embargos à complexidade do conceito, a OCDE parece utilizar *Welfare* de forma ambígua, entretanto tanto a um plexo de atributos econômico-sociais subjetivos ("prejuízos financeiros, [...] desnecessário pesar cognitiva sob o consumidor"; tradução nossa) como de preservação de direitos e garantias individuais e coletivas ("invasão de privacidade [...], redução na competitividade"; tradução nossa) (OCDE, 2020, p. 5).

## 4. A JURIDICIDADE E A DISCIPLINA DOS DARK PATTERNS NO BRASIL

A partir da análise do "status" regulatório dos *Dark Patterns* no exterior, foi possível constatar uma sólida tendência em proibi-los, o que é visto particularmente nos casos ajuizados pela FTC e na legislação dos EUA. Contudo, nos posicionamentos de entidades na União Europeia e comentários às Diretrizes 03/2022 sublinhou-se a necessidade de elucidar algumas questões referentes a estas técnicas, como o seu conceito, os limites entre um *Dark Pattern* e uma interface aceitável, se a sua proibição indiscriminada é desejável, como essa temática se relaciona com uma miríade de áreas do direito - Direitos do consumidor, proteção de dados pessoais, empresarial etc – e se o seu controle pode ser feito de maneira igual em empresas de maior ou menor porte.

Dessas constatações, têm-se um panorama global que conduz o direito brasileiro a tomar providências. A ausência de proteção adequada aos sujeitos de direito em relação aos *Dark Patterns*, tendo em vista os efeitos constatados nas pesquisas juntadas ao capítulo 2, resulta em prejuízos sócio-econômicos incalculáveis à população.

Portanto, o objetivo deste capítulo é avaliar como os *Dark Patterns* podem ser juridicamente equacionados no Brasil, o que será trabalhado comparativamente com o direito estrangeiro em tópicos dedicados a diferentes aspectos do tema. Nos subcapítulos 4.1 e 4.2 serão abordadas as pesquisas nacionais sobre os *Dark Patterns* e o seu conceito. No capítulo 4.3 discutiremos sua juridicidade e regulamentação no país.

### 4.1 AS PESQUISAS SOBRE *DARK PATTERNS* NO BRASIL

Não existem muitas pesquisas científicas brasileiras sobre a temática, tendo sido encontrados 2 estudos referentes primordialmente a ciências sociais - um artigo jurídico e um artigo de abordagem administrativa – e 3 estudos atinentes às ciências da computação - duas monografias e uma tese doutoral.[82]

---

[82] Deliberadamente excluiu-se colunas e artigos rápidos de baixo rigor metodológico, publicados em sites informais e não submetidos a crivo editorial científico.

No concernente aos trabalhos em ciências sociais, o primeiro deles foi o anteriormente mencionado "Interfaces Maliciosas: Estratégias de Coleta de Dados", da autoria de André Lemos e Daniel Marques (2019). Os pesquisadores preferiram definir trazer os *Dark Patterns* à língua vernácula como "interfaces maliciosas" (IMs), caracterizando-as como tecnologias de impacto para a proteção dos dados pessoais de titulares no contexto do que denominam "a cultura da PDPA – plataformização da sociedade [...], dataficação [...] e performatividade algorítmica [...]" (LEMOS; MARQUES, 2019, p. 2): "fenômenos que caracterizam o atual estado da cultura digital, apontando para a expansão das plataformas digitais na mediação do cotidiano" (LEMOS; MARQUES, 2019, p. 2).

O artigo apreciou as IMs em dez aplicativos "escolhidos por oferecerem serviços de interesse público na capital baiana, desenvolvidos por empresas privadas ou pelo Estado" (LEMOS; MARQUES, 2019, p. 4). Para isso, os autores desenvolveram uma "escala de gravidade" de IMs dividida em níveis "leve (1)", que recolhem dados desnecessários, mas desempenham alguma funcionalidade; "moderado (2)", que facilitam o compartilhamento dos dados relacionados ao serviço com terceiros; e "grave (3)", que simultaneamente propiciam a coleta de dados desnecessários e o compartilhamento com terceiros. (LEMOS; MARQUES, 2019, p.4-5)

Foram listadas diversas espécies de IMs em escala de gravidade predominantemente leve (70% dos encontrados), mas a quantidade desses Padrões era expressivamente maior nos aplicativos da iniciativa privada (LEMOS; MARQUES, 2019).

Em conclusão, os autores salientaram a necessidade de expandir o escopo das pesquisas a partir de "análises suplementares de documentos (de licitação do serviço, manual técnico dos aplicativos), conversa com desenvolvedores, *survey* com usuários, [e] ampliação do corpus empírico" (LEMOS; MARQUES, 2019, p. 9).

O segundo é a recente pesquisa "Pensa que me Engana, eu Finjo que Acredito: Padrões Obscuros sob a Perspectiva do Usuário" (2022), fruto do XLVI encontro da Associação Nacional de Pós-Graduação e Pesquisa em Administração (ANPAD) ocorrido entre 21 e 23 de setembro de 2022, elaborado por Luiz Calonga, Carla Soares, Thiago Melo e Luciano Machado.

Neste trabalho, os autores sugeriram a tradução do termo para Padrões Obscuros (CALONGA *et al*, 2022), empregando o modelo taxonômico de Luguri e Strahilevitz (2021) como o referencial teórico do estudo (CALONGA *et al*, 2022).

Foram realizadas "6 entrevistas - 4 entrevistas individuais e 2 grupos focais [...]" (CALONGA *et al*, 2022, p. 10) -, questionando os 11 participantes sobre suas percepções individuais sobre os *Dark Patterns*, qual é a responsabilidade e motivação por trás de seu uso, quais seriam as formas de prevenção e se eles os aceitam ou não (CALONGA *et al*, 2022).

Em síntese, os participantes afirmaram que já haviam se deparado com essas interfaces, apesar de não conhecê-las como "*Dark Patterns*", e julgaram que as empresas são as principais culpadas em sua implementação, com uma pequena parcela de culpa dos usuários em razão da ausência de pensamento crítico, constatando-se um oscilante nível de aceitação. (CALONGA *et al*, 2022, p. 18)

Assim, destes 2 primeiros estudos é possível afirmar que as contribuições das pesquisas em ciências sociais sobre os *Dark Patterns* no Brasil são bastante limitadas. Por certo, ambos os trabalhos são valiosos na medida que propuseram traduções para os *Dark Patterns* (interfaces maliciosas e padrões obscuros) e introduziram uma temática desconhecida às academias. Dito isto, os artigos lidam com o problema de forma superficial, observando, o primeiro, somente os *Dark Patterns* em amostragem diminuta de 10 aplicativos e, o segundo, a percepção subjetiva de uma parcela irrisoriamente baixa e pouco diversa da população[83,84].

Sobre os estudos em ciências da computação, os dois primeiros, de Eduardo Teixeira Carneiro (2022) e de Patrícia Raposo Santana Lima (2021) avaliaram os *Dark Patterns* no contexto, respectivamente, da coleta de *cookies*[85] em 35 site de notícia

---

[83] "As entrevistas, individuais e grupos focais, foram conduzidas em junho de 2021 e envolveram 11 pessoas, como descrito na Tabela 2. Os participantes desta pesquisa são todos brasileiros, têm entre 22 e 25 anos, cursam ou concluíram o ensino superior, têm acesso à internet e têm costume de fazer compras online. Buscou-se uma heterogeneidade na formação acadêmica, compreendendo alunos que cursam ou cursaram Engenharia de Produção, Engenharia Química, Engenharia Mecânica, Direito, Economia e Administração." (CALONGA et al, 2022, p. 10) Note-se que não foram incluídas pessoas de menor renda, grau de escolaridade e de idade avançada, circunstâncias que, como se observou em pesquisas como a de Lior e Strahilevitz (2021), causam impacto considerável sobre a percepção dos *Dark Patterns*.

[84] Não está sendo avaliada a qualidade dos artigos em apreço, que evidentemente não buscaram exaurir o assunto. Tão somente observa-se que o atual quadro de pesquisa brasileiro sobre os *Dark Patterns* é irrisório, não sendo possível estabelecer uma base dogmática sólida com apenas um par de artigos.

[85] De acordo com Felipe Palhares (2022, p. 14-15) "*Cookies* são pequenos arquivos de texto que são armazenados no terminal do usuário (cliente) e que são deixados pelo servidor *web* antes que o ciclo

mais acessados do Brasil (CARNEIRO, 2022) e da rede social Instagram. O terceiro, de Jônatas Kerr de Oliveira (2022), se dedicou ao estudo de interfaces manipulativas no âmbito dos videogames, analisando os *Dark Patterns* como um dos elementos da abordagem.

Carneiro e Lima adotaram uma estrutura bastante similar, avaliando as plataformas alvo em busca de *Dark Patterns* e linguagens manipulativas com base na taxonomia primária de Brignull (CARNEIRO, 2022) (LIMA, 2021). Ambos foram capazes de identificar variadas espécies desses Padrões, estabelecendo, no que se destacam em relação a outros trabalhos científicos, uma relação com a legislação de privacidade brasileira (CARNEIRO, 2022) (LIMA, 2021).

A pesquisa de Eduardo Carneiro foi direcionada principalmente à análise dos avisos de cookies conforme a Lei Geral de Proteção de Dados (LGPD - lei n.º 13.709/18) (CARNEIRO 2022), enquanto a de Patrícia Lima à das interfaces do Instagram de acordo com três princípios éticos: "consentimento informado, controle sobre o uso dos dados e habilidade de restringir processamento dos dados." (LIMA, 2021, p. 28). As duas monografias apresentaram pontos de intercruzamento quanto aos dispositivos violados, mormente acerca dos requisitos para o consentimento previstos no art. 5º, XII, da LGPD (BRASIL, 2018)[86] (CARNEIRO, 2022, p. 1-2 e 27-28) (LIMA, 2021, p. 60). Patrícia Lima também acrescentou a necessidade de observar o princípio do *Privacy by Design* e seus subprincípios, que seriam violados pelos *Dark Patterns* (LIMA, p. 17-21).

Já o trabalho de Carneiro trouxe duas considerações interessantes de menor expressividade na doutrina, mas que sem embargos proporcionam oferecem perspectiva relevante para a regulamentação destas técnicas.

O autor buscou diferenciar os *Dark Patterns* do que chamou de *Dark Strategies* (CARNEIRO, 2022, p. 21), que se caracterizariam pelo uso conjunto de *Dark Patterns* com outros métodos de comunicação para exercer determinados efeitos sob o consumidor (CARNEIRO, 2022, p. 21). Em outras palavras, *Dark*

---

da comunicação por meio do protocolo HTTP se encerre. Toda vez que o usuário acessa novamente o mesmo *website*, os *cookies* inicialmente armazenados são lidos pelo servidor *web*, o que permite várias funcionalidades, como lembrar os itens colocados em um carrinho de compras, autenticar o usuário para o acesso à conta, personalizar aspectos da exibição da página (tais como a língua de exibição, padrões de medida, horários) ou mesmo acompanhar o comportamento do usuário na página".

[86] LGPD, art. 5º, XII: "consentimento: manifestação livre, informada e inequívoca pela qual o titular concorda com o tratamento de seus dados pessoais para uma finalidade determinada".

*Strategies* seriam os objetivos dos agentes com o uso dessas técnicas.[87] Além disso, Carneiro reiterou a existência de *Bright Patterns* - padrões de design benéficos que devem ser incentivados (CARNEIRO, p. 28) - algo defendido em alguns dos comentários às Diretrizes 03/2022 da EDPB (UNIÃO EUROPEIA, 2022).

Por outro lado, a tese de Jônatas Kerr de Oliveira (2022) trabalhou, de forma bastante única, os problemas do conceito e dos limiares entre um *Dark Pattern* e uma prática diversa. O pesquisador identificou que o conceito do que seria uma dessas técnicas é bastante relativo, e por isso está equivocado afirmar que algumas interfaces citadas pela doutrina são sempre *Dark Patterns*. Por exemplo, o "Grinding" - a repetição de tarefas como métrica de progresso em um jogo -, descrito por Zagal *et al* (2013) como *Dark Patterns*, não necessariamente implica o objetivo de manipular os jogadores, podendo ser utilizado para "fornecer mais tempo de jogo sem necessariamente gastar mais com desenvolvimento [...]" (OLIVEIRA, 2022, p. 44).

Desta forma, Oliveira entende que cada interface precisa ser analisada segundo o contexto de sua implementação e, ainda, de acordo com os efeitos benéficos ou maléficos que produz, pois a apuração da intencionalidade dos *Dark Patterns* é dificultosa e, em alguns casos, impossível (OLIVEIRA, 2022, p. 69). Isto implica que, segundo o autor, porquanto as definições de *Dark Pattern* trazidas pela doutrina ressaltam seu caráter intencional, deixando de computar designs que acidentalmente trazem efeitos negativos, eles não seriam figuras adequadas para embasar o estudo de interfaces prejudiciais. (OLIVEIRA, 2022, p. 69)[88]

---

[87] "Existem diversas maneiras de transmitir uma informação aos usuários, desde uma forma clara e objetiva até uma maneira sorrateira e subjetiva. Seguindo a mesma lógica aplicada por Brignull (2013), quando essa comunicação utiliza de artifícios obscuros, como dark patterns e textos que expressam de forma mascarada e subjetiva a informação ao usuário, com o objetivo de influenciar o usuário, podemos considerá-la como uma dark strategy, ou seja, uma estratégia obscura." (CARNEIRO, 2022, p. 21)

[88] Em resumo, não é possível observar intencionalidade em "padrões de jogo", pois são abstrações sem contexto de jogo, e se for possível observar alguma intencionalidade, esta estaria na camada de contextualização, que contém as mecânicas e os recursos audiovisuais de um jogo. [...] Como a abordagem dos Dark Patterns faz menção à intenção do designer como parâmetro definidor do conceito, esse cenário seria mais um empecilho à sua utilização para os propósitos delimitados especificamente por esta tese. [...] Considerando esses problemas conceituais, não podemos recorrer ao conceito de Dark Patterns para identificar um dado padrão como intencionalmente feito para coagir ou manipular um jogador. [...] (OLIVEIRA, 2022, p. 69)

## 4.2 O CONCEITO DE *DARK PATTERN*

Como muitas vezes reforçado, a doutrina e as normas jurídicas conceituaram os *Dark Patterns* de múltiplas formas, não sendo possível acordar sobre qual seria a integralidade de elementos que os definem. Para visualizar e compreender esta pluralidade conceitual, na tabela a seguir listamos, não exaustivamente, alguns dos principais conceitos (as descrições dessas técnicas) e nomenclaturas (o nome conferido pela doutrina a estas técnicas, como *Dark Patterns* ou Interfaces Maliciosas) atribuídos a estas técnicas:

Tabela 4 - Definições de Dark Patterns

| Número | Nomenclatura | Conceito | No idioma original | Fonte |
|---|---|---|---|---|
| 1 | *Dark Patterns.* | - ("Atualmente, não existe um conceito legal para o que são *"Dark Patterns"*") | - ("*There is currently no legal concept of what "Dark Patterns" are.* [...]") | BEUC, 2022, p. 5 |
| 2 | *Dark Patterns.* | "Um dark pattern é uma interface cuidadosamente construída para enganar usuários a fazer coisas que normalmente não fariam". (tradução nossa) | *"A dark pattern is a user interface carefully crafted to trick users into doing things they might not otherwise do"* | BRIGNULL, 2013. |
| 3 | Padrões Sombrios; Padrões Obscuros. (*Dark Patterns*) | "é um design de interface compreendido como enganoso, que induz usuários a executarem ações não intencionais e indesejadas". | *idem* | CALONGA *et al*, 2022, p. 3. |
| 4 | *Dark Patterns.* | *"Dark Patterns* são funcionalidades de design usadas para enganar, dirigir, ou manipular usuários para adotarem comportamentos rentáveis a um serviço online, mas que são recorrentemente prejudiciais aos usuários ou contrários aos seus interesses." (tradução nossa) | *"Dark Patterns are design features used to deceive, steer, or manipulate users into behavior that is profitable for an online service, but often harmful to users or contrary to their intent"* | CHOPRA, 2020, p.1. |

| 5 | Interfaces maliciosas. (*Malicious interfaces*) | Interfaces maliciosas são designs construídos com a intenção de deliberadamente sacrificar a experiência de usuário em uma tentativa de alcançar os objetivos do designer à frente dos do usuário. | "*The key difference between usable interface design and malicious interface design is the intent on the part of the designer to deliberately sacrifice the user experience in an attempt to achieve the designer's goals ahead of those of the user.*" | CONTI; SOBIESK, p. 271. |
|---|---|---|---|---|
| 6 | *Dark Patterns.* | "No contexto destas Diretrizes, "*Dark Patterns*" são consideradas interfaces e experiências de usuário implementadas em plataformas de mídia social que levam os usuários a tomar decisões não intencionais e potencialmente prejudiciais acerca do processamento de seus dados pessoais." (tradução nossa) | "*In the context of these Guidelines, "Dark Patterns" are considered as interfaces and user experiences implemented on social media platforms that lead users into making unintended, unwilling and potentially harmful decisions regarding the processing of their personal data.*" | EDPB, p. 2. |
| 7 | *Dark Patterns.* | "Instâncias onde os designers utilizam seus conhecimentos sobre o comportamento humano (p.e., psicologia) e os desejo dos usuários-fim para implementar funcionalidades enganosas que não são do melhor interesse do usuário." (tradução nossa) | "*We use the term dark patterns to define instances where designers use their knowledge of human behavior (e.g., psychology) and the desires of end users to implement deceptive functionality that is not in the user's best interest.*" | GRAY *et al.*, 2018, p. 1. |
| 8 | *Dark Patterns.* | "*Dark Patterns* são interfaces de usuário cujos designers sabidamente confundem os usuários, dificultando a expressão das reais preferências dos usuários, ou manipulando os usuários a tomar certas ações." (tradução nossa) | "*Dark Patterns are user interfaces whose designers knowingly confuse users, make it difficult for users to express their actual preferences, or manipulate users into taking certain actions.*" | LUGURI; STRAHILE VITZ, 2021, p. 44. |
| 9 | *Dark Patterns.* | "*Dark Patterns* são escolhas de design de interface de usuário | "*Dark Patterns are user interface design choices that benefit an online service by*" | MATHUR *et al.*, 2019, p. 2. |

| | | que beneficiam um serviço online por coagir, dirigir ou enganar usuários a tomar decisões que, se completamente informadas e capazes de selecionar alternativas, eles poderiam não tomar." (tradução nossa) | *coercing, steering, or deceiving users into making decisions that, if fully informed and capable of selecting alternatives, they might not make."* | |
|---|---|---|---|---|
| 10 | *Dark Patterns.* | "Estratégias que são utilizadas em sítios eletrônicos e aplicações que obrigam as pessoas a fazerem coisas que não desejam, como comprar ou registrar-se em algo." (tradução nossa) | *"estrategias que se utilizan en los sitios web y aplicaciones que obligan a las personas a hacer cosas que no quieren, como comprar o registrarse en algo."* | SERNAC, 2021, p. 3. |
| 11 | *Dark Patterns.* | Design de interfaces ou funcionalidades que sutilmente manipulam pessoas a tomarem decisões subótimas. (tradução nossa) | *"Design interfaces or features that subtly manipulate people in making suboptimal decisions".* | SIN *et al.,* 2021, p. 1. |
| 12 | *Dark Patterns* de design de jogos (*Dark game design Pattern*, em tradução livre). | "Um *Dark Pattern* de design de jogos é um padrão usado intencionalmente pelo criador do jogo para causar experiências negativas contrárias aos melhores interesses dos jogadores e sem o seu consentimento." (tradução nossa | *"A dark game design pattern is a pattern used intentionally by a game creator to cause negative experiences for players that are against their best interests and happen without their consent."* | ZAGAL; BJÖRK; LEWIS, 2013, p. 3. |
| 13 | - | "Interface de usuário com o propósito ou substantivo efeito de obscurecer, subverter, ou limitar a autonomia, capacidade decisória, ou escolha dos usuários para obter consentimento ou dados." (tradução nossa) | *"[...] user interface with the purpose or substantial effect of obscuring, subverting, or impairing user autonomy, decision-making, or choice to obtain consent or user data."* | DETOUR act (Sec. 3, (a), (1)). |
| 14 | *Dark Patterns.* | ""Dark pattern" significa interface de usuário desenhada e manipulada com o | *""Dark pattern" means a user interface designed or manipulated with the substantial effect of* | CCPA/CP RA (cal.civ.cod e, |

| | | substancial efeito de subverter ou limitar a autonomia, liberdade de decidir ou de escolher do usuário, conforme definido em lei." (tradução nossa) | *subverting or impairing user autonomy, decisionmaking, or choice, as further defined by regulation."* | §1798.140, I). |
|---|---|---|---|---|
| 15 | *Dark Patterns.* | ""Dark pattern" significa interface de usuário desenhada e manipulada com o substancial efeito de subverter ou limitar a autonomia, liberdade de decidir ou de escolher do usuário." (tradução nossa) | *""Dark pattern" means a user interface designed or manipulated with the substantial effect of subverting or impairing user autonomy, decision-making, or choice."* | CPA (Col. Rev. Stat., §6-1-1303, (9)). |
| 16 | *Dark Patterns* (vide capítulo 3) | "[...] Interfaces online [...] que enganam, manipulam ou, de outra forma, materialmente distorce ou limita a habilidade dos receptores do serviço de fazer decisões livres e informadas" (tradução nossa) | *"[...] online interfaces [...] that deceives, manipulates or otherwise materially distorts or impairs the ability of recipients of their service to make free and informed decisions."* | DSA, article 23a, (1). |

Fonte: criado pelo autor

A partir da análise da listagem, fica evidente que o termo preferido pelos pesquisadores e legisladores é *Dark Patterns*, em proporção de 15 contra 1, devendo ser adotado para evitar dissonâncias acadêmicas. Em português, a tradução Padrão Obscuro proposta por CALONGA *et al.* (2022) é apropriada, guardando o maior grau de semelhança com o nome original.

Por outro lado, as definições somam maior grau de complexidade, demandando uma análise mais aprofundada. Em um primeiro momento, verifica-se a repetição dos resultados da pesquisa de Mathur *et al.* (2021) sobre as múltiplas definições para "*Dark Pattern*", extraindo-se quatro observações: (i) atribuem-se inúmeras características às interfaces, como "truques", "enganosas", "coercitivas", etc.; (ii) mencionam-se inúmeros mecanismos de influência sob os usuários, como "manipulam", "enganam", "atacam", "exploram" etc.; (iii) situam os designers responsáveis pelos *Dark Patterns* de várias maneiras na relação, abusando de seus conhecimentos sobre o comportamento humano ou intencionalmente usando *Dark Patterns* para alcançar um objetivo, por exemplo; e (iv) mencionam diferentes resultados, como prejuízos aos usuários ou o benefício da empresa (MATHUR *et al.*, 2021, p. 3-5; tradução nossa).

Primeiramente nos dirigiremos às observações de número i e ii - a excessividade de características e mecanismos empregados. Usar esses termos para dizer que as interfaces são "enganosas", ou afirmar que elas "atacam" o usuário, é circunstancialmente útil para visualizar determinados *Dark Patterns*, o que acarreta, contudo, algumas dissonâncias.

Na esteira de Mathur *et al.* (2021), algumas características podem ser encontradas consistentemente em alguns tipos de *Dark Patterns*, não implicando que todos eles as apresentarão. Os *Dark Patterns* de "culpabilização" (*confirmshamming*"), por exemplo, não são inerentemente "enganosos" ou "coercitivos", sendo melhor definidos como "manipuladores". Cada interface incorpora elementos e objetivos diferentes, projetando efeitos diversos sob os usuários em cada contexto, de maneira que inserir todos eles no conceito de *Dark Pattern* parece ser inviável – especialmente ao considerar a efemeridade do *status quo* tecnológico, nada impedindo o surgimento de interfaces que apropriem novos mecanismos, para os quais nenhum dos adjetivos prévios é suficiente. Portanto, a excessiva adjetivação do conceito de *Dark Patterns* pode causar, em sentido contrário do pretendido, a sua descaracterização.

As observações iii e iv merecem, cada uma, considerações distintas, mas de profundo impacto na definição de *Dark Patterns*. É necessário ponderar que os Padrões usados na venda de bens ou serviços online entre consumidor e fornecedor (arts. 2º e 3º do CDC) atraem a legislação consumerista para disciplinar o caso. Da mesma forma, quando eles são utilizados durante uma etapa do tratamento de dados pessoais, a LGPD deverá ser aplicada, quando não incidir alguma das hipóteses de exclusão prevista no seu art. 4º[89,90].

---

[89] LGPD, Art. 4º: "Esta Lei não se aplica ao tratamento de dados pessoais: I - realizado por pessoa natural para fins exclusivamente particulares e não econômicos; II - realizado para fins exclusivamente: a) jornalístico e artísticos; ou b) acadêmicos, aplicando-se a esta hipótese os arts. 7º e 11 desta Lei; III - realizado para fins exclusivos de: a) segurança pública; b) defesa nacional; c) segurança do Estado; ou d) atividades de investigação e repressão de infrações penais; ou IV - provenientes de fora do território nacional e que não sejam objeto de comunicação, uso compartilhado de dados com agentes de tratamento brasileiros ou objeto de transferência internacional de dados com outro país que não o de proveniência, desde que o país de proveniência proporcione grau de proteção de dados pessoais adequado ao previsto nesta Lei. [...]".
[90] No entanto, a incidência das hipóteses de exceção nas relações envolvendo *Dark Patterns* é rara, pois estas em sua maioria se desenvolvem num plano comercial que foge ao uso exclusivamente particular, jornalístico, artístico ou acadêmico. Na mesma senda, como observado em todos os casos

Neste sentido, os *Dark Patterns* devem ser analisados conforme a natureza do objeto jurídico tutelado e os objetivos das normas aplicáveis: a proteção do Consumidor e do Titular de Dados Pessoais.[91] Ambos os sujeitos gozam, respectivamente, de presunção de vulnerabilidade perante os fornecedores, por força do art. 4º, I, do CDC, e os agentes de tratamento, a partir de construção doutrinária explicitada no trabalho de Karoline Silveira Marculino (2021). Isto vai ao encontro da demanda de comentaristas às Diretrizes 03/2022 pelo reconhecimento da vulnerabilidade dos usuários expostos aos *Dark Patterns* (UNIÃO EUROPEIA, 2022).

Decorrendo disso, diante da responsabilidade civil da empresa ou agente pelos danos oriundos do uso de *Dark Patterns*, Instrumento singular de proteção dos sujeitos tutelados, não há como encaixar o designer de forma juridicamente relevante nas relações de consumo ou de tratamento de dados. O CDC é categórico ao fixar a responsabilidade da empresa pelo fornecimento inadequado de seus produtos ou serviços (arts. 12 e 14 do Código), vedada a denunciação da lide para condenar outros responsáveis (art. 88 do CDC). Similarmente, a LGPD estabelece, em seu art. 42, a responsabilidade solidária dos agentes envolvidos no tratamento de dados pessoais.

Desta feita, a proteção dos direitos do consumidor e do titular em relação ao uso de *Dark Patterns* tem como pertinência somente a responsabilidade da empresa ou agente de tratamento, sendo desimportante a atuação de terceiros sob os cuidados desses responsabilizados.

Quanto à observação iv, uma vez posicionada a defesa dos vulneráveis como pilar dos sistemas de defesa do consumidor e da proteção de dados pessoais, os mais importantes resultados da implementação de um *Dark Pattern* parecem ser os

---

práticos mencionados nesta tese, as relações nas quais aplicam-se os *Dark Patterns* para a coleta de dados pessoais dizem respeito principalmente a relações privadas, sem atuação de interesse público do Estado que faça incidir o art. 4º, III, da LGPD. Por último, se os *Dark Patterns* de uma empresa estrangeira alcançaram o titular brasileiro através da internet não há como aplicar o art. 4º, IV, da LGPD, porque configurada uma transferência internacional de dados, nos termos da definição do art. 5º, XV, da LGPD ("transferência de dados pessoais para país estrangeiro ou organismo internacional do qual o país seja membro").

[91] O CDC afirma, em seu art. 4º, caput, que a Política Nacional das Relações de Consumo visa a proteção dos direitos e garantias fundamentais dos consumidores, assegurando-lhes o "atendimento de suas necessidades e o respeito à sua dignidade, saúde e segurança, a proteção de seus interesses econômicos, a melhoria da sua qualidade de vida, bem como a transparência e harmonia das relações de consumo", o que reflete na atual conjuntura dos princípios e direitos de proteção ao consumidor previstos nos arts. 4º e 6º. Por sua vez, a LGPD, no art. 1º, erige como seu máximo objetivo "proteger os direitos fundamentais de liberdade e de privacidade e o livre desenvolvimento da personalidade da pessoa natural", o que é ratificado pelo art. 17 da norma.

danos ao patrimônio e a violação de direitos dos usuários. Não é dizer que o enriquecimento ilícito da empresa não possua relevância, mas o prejuízo aos consumidores e titulares é o elemento essencial ante os parâmetros jurídicos em jogo. Em outras palavras, um *Dark Pattern* não deve necessariamente proporcionar lucro ou vantagem ao agente para que seja alvo de controle pelo ordenamento, bastando a capacidade de lesar os usuários.

Além das quatro facetas identificadas por Mathur *et al.* (2021), outro elemento de recorrente[92] associação aos *Dark Patterns*, como destacado por Oliveira (2022), é a intencionalidade, o que é visto claramente nas entradas de número 2, 5 e 8 na tabela.

Rememorando o sustentado pelo autor, a intencionalidade atribuída aos *Dark Patterns* dificultaria a adoção do conceito para definir técnicas prejudiciais em design de jogos, pois algumas interfaces podem prejudicar a experiência do usuário de forma não pretendida originalmente (OLIVEIRA, 2022). Para o direito, este argumento é apropriável sob a ótica do regime de responsabilidade aplicável aos fornecedores e agentes de tratamento (arts. 12 e 14 do CDC; art. 42 da LGPD).

Em uma exagerada simplificação, são dois os principais regimes de responsabilidade civil em nosso ordenamento jurídico: subjetivo e objetivo. Os elementos da responsabilidade subjetiva ou aquiliana, de acordo com Flávio Tartuce (2021), são a conduta humana, a culpa, o nexo de causalidade e o dano (TARTUCE, 2021, p. 464). Porém, o regime de responsabilidade objetiva aplicável aos entes que utilizam os *Dark Patterns* em produtos, serviços e tratamentos de dados é objetiva, a qual exclui como requisito a culpa dos agentes para a atribuição dos deveres reparatórios (TARTUCE, 2021, p. 521-529). Portanto, sob esta visão, mesmo que possa consistir propriedade importante do estudo dessas técnicas sob o escopo do UX/UI e da ética profissional dos designers - hipóteses fora do escopo deste trabalho -, a intencionalidade não parece ser apropriada para integrar o conceito jurídico de *Dark Pattern*, vez que se trata de característica inserida no elemento subjetivo "culpa".

Adicionalmente, caso se aceitasse a inclusão do atributo não obstante a responsabilidade da empresa, estaríamos concordando com a possibilidade da produção de danos aos usuários não ser vista com a precisa seriedade, além do que a intencionalidade é elemento de difícil e inconsistente apuração (OLIVEIRA, 2022), situação que acarreta, ulteriormente, a impunidade desses agentes.

---

[92] Contudo, não é elemento presente em todos os conceitos, visto que as definições do CCPA/CPRA e CPA (entradas 14 e 15 da tabela) listam como elemento substancial para a caracterização de um *Dark Pattern* a produção de efeitos, despicienda a intencionalidade.

Os resultados destas análises nos conduzem a propor o seguinte conceito jurídico: *Dark Pattern* consiste em toda interface capaz de substantivamente violar direitos e causar danos à pessoa natural.

A definição é bastante distinta das outras propostas doutrinárias, especialmente aquelas que visam descrever ao máximo as características do fenômeno. Sob esta ótica seria possível, *a priori*, acusar o novo conceito de ser demasiadamente simples ou desconexo com as propriedades dos *Dark Patterns*.

Contudo, apreciando as dissonâncias conceituais encontradas ao longo deste subcapítulo, a definição apresentada é a mais compatível com o ordenamento jurídico nacional. Acerca dos pontos i e ii de Mathur *et al.* (2021), ela elimina a necessidade de listar os múltiplos mecanismos e efeitos mencionados pela doutrina, que não dizem respeito univocamente a todos os *Dark Patterns*. Considerar a pluralidade de tipos de *Dark Pattern* não somente dificultaria encontrar uma definição adequada como a enrijeceria, prendendo-a a técnicas que poderiam cair em desuso ou ser substituídas com o tempo. Assim, o conceito proposto contorna estes obstáculos ao minimizar os seus componentes a 3:

a) Interface: o dicionário Priberam conceitua interface como "meio através do qual um usuário interage com um sistema operacional ou com um programa" (2021). Trata-se do elemento técnico que compõe os *Dark Patterns*, conferindo-lhe identidade e maleabilidade para se adequar com as tecnologias de interação humano-computador (HCI) atuais (recursos gráficos, imagens, textos etc.) e vindouras, bem como com qualquer tipo de *Dark Pattern* existente, quaisquer sejam suas características. Neste sentido, considerando que textos e métodos de comunicação são elemento da interface e a intencionalidade não comporta espaço nesta definição, não concebemos como juridicamente relevante a distinção entre *Dark Pattern* e *Dark Strategy* sustentada por Carneiro (2022).

b) Violação de direitos e danos substantivos: os resultados de primário interesse jurídico da implementação dos *Dark Patterns*, na perspectiva da proteção dos usuários.

c) Pessoa natural: os indivíduos tutelados em face dos danos causados pelos *Dark Patterns*. A possibilidade de uma pessoa jurídica sofrer danos oriundos da aplicação de um *Dark Pattern* é discutível, mas seria substantivamente limitada se comparada à do indivíduo vulnerável inserido em uma relação de consumo ou de tratamento de dados pessoais. Os resultados de pesquisas como as de Luguri e Strahilevitz (2021), Blake *et al.* (2018) e Sin *et al.* (2022) apontam para a alta influência de Dark Patterns sob os consumidores, deduzindo-se a existência de um considerável déficit técnico que reflete na precariedade de suas escolhas, ao passo que as empresas estariam melhor equipadas para evitar os seus efeitos, através de um corpo de funcionários e meios organizacionais para otimizar o processo decisório. Assim, optou-se por limitar a tutela contra *Dark Patterns* apenas à pessoa natural.

Não há relevância em trazer a figura do designer da interface para disciplinar uma relação travada entre um indivíduo e uma empresa, de acordo com os fundamentos trazidos acima, e por isso ela também foi excluída do conceito. Outrossim, em razão do regime de responsabilidade objetiva pelo qual respondem as empresas em questão e o objetivo principal de resguardar direitos e garantias individuais, retirou-se as menções à intencionalidade e necessidade de acarretar benefício ao implementador do *Dark Pattern*.

A definição proposta não retira do ordenamento jurídico outras características que os *Dark Patterns* possam deter. O conceito foi desenvolvido com a função de abstrair destas técnicas elementos inconsistentes ou não necessariamente relevantes para caracterizar a problemática no direito brasileiro. Ele não pretende, desta forma, exaurir o conteúdo material dos *Dark Patterns*, que compreendem interações e dinâmicas sociais de considerável complexidade.

## 4.3 A ANTIJURIDICIDADE DOS *DARK PATTERNS*

Constatando-se o potencial lesivo dos *Dark Patterns* nos estudos e experiências dissecados ao longo desta tese, desponta uma forte tendência global das normas de proteção de dados e defesa do consumidor para proibi-los, o que se faz evidente nas opções legislativas tomadas nos Estados Unidos, a partir de normas como DETOUR, CCPA, CPRA e CPA, e na União Europeia, através do recentemente aprovado DSA. Diante disso, cabe averiguar se estas respostas são adequadas para o ordenamento jurídico nacional, levando em conta as características dos *Dark Patterns* e as normas de proteção de dados e direito do consumidor preexistentes.

### 4.3.1 Os *Dark Patterns* segundo a proteção de dados pessoais

Os *Dark Patterns* interferem substancialmente no poder decisório dos usuários sobre o uso de seus dados pessoais, e por isso aparentemente se demonstram contrários aos próprios fundamentos dos institutos modernos da proteção de dados: o livre desenvolvimento da personalidade através da autodeterminação informativa.

Em sua origem histórica, tomando como base as preocupações com o abuso de meios tecnológicos para a captação de dados e, posteriormente, a guarda da privacidade dos indivíduos, a proteção de dados pessoais foi representada pela lei como um direito de caráter negativo em face do Estado, o que perdurou durante a 1ª e 2ª geração dessas normas (DONEDA, 2021, p. 180-181).

Todavia, não se demorou em reconhecer que imbuir esse direito apenas com caráter excludente não é bastante para equacionar o uso de dados pelo poder público[93], vista a impossibilidade de negar-lhe o tratamento diante da instrumentalidade dos dados pessoais para o exercício das funções estatais. (DONEDA, 2021, p. 181-182)

Buscou-se, assim, um novo paradigma para enxergar a dinâmica da proteção de dados pessoais: a autodeterminação informativa. (DONEDA, 2021, p. 181-182) Este elemento foi introduzido com maior força em importante decisão da Corte Constitucional germânica sobre a Lei do Censo Alemã de 1983, a qual, declarando a parcial inconstitucionalidade da norma em virtude da ampla e desmesurada captura dos dados pessoais de cidadãos, firmou o entendimento de que a "capacidade do indivíduo de autodeterminar seus dados pessoais seria parcela fundamental do seu direito em livremente desenvolver sua personalidade". (BIONI, 2021, p. 99)

---

[93] Sobre as normas de primeira geração: "estas leis de proteção de dados de primeira geração não demoraram muito a se tornarem ultrapassadas, diante da multiplicação dos centros de processamento de dados, o que dificultou propor um controle baseado em um regime de autorizações, rígido e detalhado, que demandava um minucioso acompanhamento. [...]" (DONEDA, 2021, p. 181). Sobre as de segunda geração: "estas leis apresentavam igualmente seus problemas, o que motivou uma subsequente mudança de paradigma: percebeu-se que o fornecimento de dados pessoais pelos cidadãos tinha se tornado um requisito indispensável para a sua efetiva participação social. Tanto o Estado como os entes privados utilizavam intensamente o fluxo de informações pessoais para seu funcionamento, e a interrupção ou mesmo o questionamento deste fluxo pelo cidadão - ou seja, a atuação direta da liberdade do cidadão de interromper o fluxo de informações pessoais - implica não raro na sua exclusão de algum aspecto da vida social. [...] Enfim, percebia-se que o exercício puramente individual desta liberdade envolvia consequências bem maiores que aquelas que diziam respeito somente às informações pessoais e eram fundamentais para a própria socialização de cada pessoa" (DONEDA, 2021. p. 182)

Nestes termos, a 3ª geração de leis passa a se desenvolver em torno da promoção da liberdade dos titulares em decidir como usar os seus dados pessoais, concretizando um direito complexo de facetas negativas e positivas individuais, o que veio a se estender à tutela coletiva na 4ª geração. (DONEDA, 2021, p. 183-184)

Consequentemente, o uso de interfaces para cercear indevidamente o poder decisório dos titulares acerca de seus dados viola a autodeterminação informativa, disposta no art. 2º, II, da LGPD, desdobrando-se no esboroamento sistemático de variados princípios e regras encontrados no art. 6º da mesma norma.

Os princípios da finalidade[94] e adequação[95] (art. 6º, I e II, da LGPD), por exemplo, pelos quais o tratamento de dados deve possuir uma finalidade legítima e informada ao titular e os métodos de tratamento devem ser adequados com a referida finalidade, seriam automaticamente violados por um *Dark Pattern* do tipo "informação oculta" - que ocultasse essas informações dos titulares para facilitar a coleta dos dados pessoais -, o que acaba por também violar o princípio da transparência[96] (art. 6º, VI, da LGPD).

Similarmente, havendo um recolhimento desproporcional de dados considerando a finalidade do tratamento, como um *Dark Pattern* que incita um usuário a compartilhar mais dados que o razoavelmente necessário, há uma violação ao princípio da necessidade[97] (art. 6º, III, da LGPD).

No caso de um *Dark Pattern* de "hotel de baratas", onde o usuário não é capaz de cancelar um tratamento de dados pessoais, a interface pode restringir inadequadamente o acesso do indivíduo aos seus dados pessoais ou a capacidade de corrigi-los ou alterá-los, ferindo os princípios do livre acesso[98] e da qualidade[99] (art. 6º, IV e V, da LGPD).

---

[94] LGPD, art. 6º, I: "finalidade: realização do tratamento para propósitos legítimos, específicos, explícitos e informados ao titular, sem possibilidade de tratamento posterior de forma incompatível com essas finalidades".

[95] LGPD, art. 6º, II: "adequação: compatibilidade do tratamento com as finalidades informadas ao titular, de acordo com o contexto do tratamento".

[96] LGPD, art. 6º, VI: "transparência: garantia, aos titulares, de informações claras, precisas e facilmente acessíveis sobre a realização do tratamento e os respectivos agentes de tratamento, observados os segredos comercial e industrial".

[97] LGPD, art. 6º, III: "necessidade: limitação do tratamento ao mínimo necessário para a realização de suas finalidades, com abrangência dos dados pertinentes, proporcionais e não excessivos em relação às finalidades do tratamento de dados".

Outro aspecto de bastante relevância ao tratar dos *Dark Patterns* na proteção de dados pessoais é o seu impacto aos requisitos do consentimento, o que foi também observado pelas legislações estrangeiras que os disciplinam, destacadamente o DETOUR e o Código Civil da Califórnia, conforme observado no capítulo 3.

A LGPD (art. 5º, XII) estabelece 3 características para a validade do consentimento (BIONI; LUCIANO, 2021, p. 153-154), as quais podemos confrontar, a critério exemplificativo, com 3 tipos de *Dark Patterns*: (i) o consentimento deve ser informado, instruindo-se o titular sobre as características do tratamento de dados pessoais para o qual se solicitou o aceite, o que é obstado por *Dark Patterns* de ocultação de informações; (ii) o consentimento deve ser livre, despido da influência desmesurada de terceiros, o que pode ser ferido por um *Dark Pattern* do tipo "registro forçado"; (iii) e o consentimento deve ser inequívoco, sem qualquer ambiguidade, algo diametralmente oposto à situação causada pelo *Dark Pattern* de tipo "pergunta capciosa". No tratamento de dados sensíveis[100], o consentimento deverá ser, ainda, específico e destacado[101], algo também facilmente contornável por estes *Dark Patterns*.

Nesse andar, se o consentimento foi a base legal escolhida para conferir legitimidade ao tratamento de dados pessoais, de acordo com as hipóteses dos arts. 7 e 11 da LGPD, o uso de um *Dark Pattern* que interfere nos seus requisitos essenciais acarretará a ausência de lastro para realizar as operações - lançando o tratamento à ilegalidade.

Por fim, alguns *Dark Patterns*, como os do tipo "hotel de baratas", medida que restringe a capacidade dos usuários de interagir com a plataforma para satisfazer suas pretensões legítimas, tem o condão de suprimir os direitos do titular previstos no art. 18 da LGPD, especialmente o direito de revogar o consentimento para o tratamento de dados (Art. 18, IX, da LGPD).

---

[98] LGPD, art. 6º, IV: "livre acesso: garantia, aos titulares, de consulta facilitada e gratuita sobre a forma e a duração do tratamento, bem como sobre a integralidade de seus dados pessoais".

[99] LGPD, art. 6º, V: "qualidade dos dados: garantia, aos titulares, de exatidão, clareza, relevância e atualização dos dados, de acordo com a necessidade e para o cumprimento da finalidade de seu tratamento".

[100] de acordo com a LGPD, os dados pessoais sensíveis são aqueles taxativamente inseridos em seu art. 5º, II: "dado pessoal sobre origem racial ou étnica, convicção religiosa, opinião política, filiação a sindicato ou a organização de caráter religioso, filosófico ou político, dado referente à saúde ou à vida sexual, dado genético ou biométrico, quando vinculado a uma pessoa natural".

[101] LGPD, art. 11, I: "quando o titular ou seu responsável legal consentir, de forma específica e destacada, para finalidades específicas"

Esta brevíssima abordagem sobre os *Dark Patterns* perante as normas proteção de dados nacionais, mal arranhando a superfície dos seus institutos, revela que a aplicação dessas técnicas fere diversos princípios e regras LGPD, evidenciando a sua antijuridicidade diante desta parcela do ordenamento brasileiro.

### 4.3.2 Os *Dark Patterns* segundo o direito do consumidor

Os efeitos dos *Dark Patterns* segundo o direito do consumidor se verificam, primeiramente, no cumprimento dos direitos básicos do consumidor, previstos no art. 6º, e incisos, do CDC, imediatamente despontando o direito básico à "educação e divulgação sobre o consumo adequado dos produtos e serviços, asseguradas a liberdade de escolha e a igualdade nas contratações" (art. 6º, II, do CDC).

De acordo com Humberto Theodoro Júnior (2021, p. 48-49), o dever de educar o consumidor se subdividiria em duas facetas: a educação formal, ministrada em instituições de ensino, e a educação informal, de encargo dos fornecedores de serviços. Desta forma, o grau de influência dos *Dark Patterns* na educação do consumidor é primariamente indireto, considerando sua capacidade de distorcer a percepção dos usuários sobre as boas práticas de consumo adotadas na internet - banalizando o uso dessas interfaces.

O maior impacto dos *Dark Patterns* ao direito à informação é verificado no tocante à segunda parte do enunciado, acerca da "liberdade de escolha e igualdade nas contratações", obrigando o "fornecedor a informar o consumidor previamente sobre as condições contratuais, evitando-se que seja surpreendido por alguma cláusula abusiva" (JÚNIOR, 2021, p. 49). O dispositivo em comento, entretanto, pois sustentado pelos princípios de "liberdade de ação e escolha da Constituição Federal (arts. 1º, III, 3º, I, 5º, caput, entre outros)" (NUNES, 2021, p. 61), não deve ser mirado exclusivamente sob o prisma contratual, expandindo-se até onde a tutela do consumidor for necessária, em harmonia com os interesses legítimos dos participantes na relação de consumo (art. 4º, III, do CDC). Desta forma, considerando os resultados largamente desfavoráveis aos consumidores advindos do uso dos *Dark Patterns*, é sustentável que essas técnicas violam diretamente a liberdade de escolha do consumidor sob as perspectivas contratual, a partir da manipulação ou ocultação dos termos da aquisição do produto ou serviço em face do consumidor, e extracontratual, trazendo danos psíquicos[102] e materiais aos consumidores.

Segunda e terceira violações a direito básico do consumidor vêm à prestação de informações adequadas sobre o produto ou serviço (art. 6º, III, do CDC), fundamentalmente minado pelos *Dark Patterns* que ocultam informações sobre o serviço prestado, e à vedação a "publicidade enganosa e abusiva, métodos comerciais coercitivos ou desleais" e "práticas e cláusulas abusivas ou impostas no fornecimento de produtos e serviços", (art. 6º, IV, do CDC), sem prejuízo da possibilidade de malferimento a outros direitos, conforme as espécies de *Dark Patterns* empregadas e seus propósitos.

O instituto da oferta previsto no CDC também é uma preocupação ao se lidar com estas técnicas, que interferem substancialmente no modo do oferecimento de produtos e serviços, como no caso FTC v. Credit Karma. O CDC, em seu art. 31, dispõe que a oferta deve ser constituída de "informações corretas, claras, precisas, ostensivas e em língua portuguesa sobre suas características, qualidades" e outros atributos. Sergio Cavalieri Filho complementa, sustentando que a oferta deve ser revestida de seriedade, "na medida em que deve ser crível, na qual o consumidor possa confiar"; completude, "na medida em que deve conter todos os elementos que vão integrar o contrato, de modo a permitir a simples adesão do consumidor"; e receptividade, "ou seja, passível de aceitação por consumidores identificados ou identificáveis", sendo ainda fundada na transparência (FILHO, 2022, p. 190-191).

Neste giro, os *Dark Patterns* podem macular oferta em razão de fomentarem a ausência de clareza, precisão e ostensividade sobre as suas características, comprometendo sua seriedade, transparência e, em alguns casos, a completude e a receptividade, nas hipóteses de *Dark Patterns* de ocultação de preços e valores até o final do processo de compra. Por isso, a oferta de bens através de uma plataforma que empregue *Dark Patterns* nesta etapa da aquisição de produtos e serviços é inválida, atraindo a aplicação do art. 35 do CDC e o aparato sancionatório consumerista para remediar a situação.

Curiosamente, a jurisprudência lidou com situações em que o emprego de interfaces comprometeu o dever de informar ostensivamente, apesar de não mencionar os *Dark Patterns*. Nos Embargos de Declaração no Recurso Especial n.º 1737428/RS, de relatoria da Ministra Nancy Andrighi, denegou-se o pedido para alterar o acórdão que condenou a empresa embargante, Ingresso Rápido, por falhar com o dever de informar. Dos autos, extrai-se que a prática combatida consiste na ocultação de taxas de conveniência até o final da aquisição do ingresso virtual, interface que se encaixa precisamente nos moldes de um *Dark Pattern* de "preços ocultos".

---

[102] Conforme determinado por Luguri e Strahilevitz (2021), observados impactos psicológicos negativos aos consumidores conforme o grau de *Dark Patterns* utilizado, e por Conti e Sobieski (2010), a despeito do baixo número de participantes na pesquisa.

Desta feita, a partir de uma análise sucinta do CDC e da doutrina consumerista os *Dark Patterns* também parecem violar diversos dispositivos referentes à proteção do consumidor, sendo possível sustentar, por este ângulo, a sua antijuridicidade.

### 4.3.3 Considerações finais sobre a juridicidade dos *Dark Patterns* e sua regulamentação

Não obstante a constatação dos fortes indícios da antijuridicidade dos *Dark Patterns* segundo a proteção de dados pessoais e o direito do consumidor, algumas situações levam a questionar se isso é de fato o que ocorre, bem como qual seria a melhor maneira de abordá-los em uma perspectiva normativa. Os EUA e a União Europeia foram duros em banir o uso dos Padrões através de suas mais novas leis - nos Estados Unidos, DETOUR Act e Safe Data Act (estes ainda em trâmite legislativo), CCPA/CPRA e CPA, e na União Europeia o DSA - estritamente proibindo a sua implementação (DETOUR, Safe Data e DSA) ou negando a sua aplicação para a coleta do consentimento (CCPA/CPRA e CPA).

Estas normas possuem acertado fundamento nos perigos advindos da capacidade dos *Dark Patterns* de influenciar os usuários, causando danos e violações a direitos em parte significativa dos casos em que são utilizados. Entretanto, a regulamentação destas técnicas não se atentou a elementos de elevada complexidade que precisam ser equacionados previamente à sua proibição.

O primeiro a ser destacado é a existência de *Dark Patterns* aplicados em diferentes condições, com potenciais de manipulação e impactos aos usuários bastante distintos um do outro, o que foi abordado por Luguri e Strahilevitz (2021). A partir de seu estudo empírico, de acordo com o abordado no segundo capítulo, *Dark Patterns* configurados de maneira "agressiva" exercem um efeito manipulativo superior sobre os usuários, às custas de uma percepção negativa e um impacto psicológico maior (LUGURI; STRAHILEVITZ, 2021). Por outro lado, os *Dark Patterns* em condição "suave" possuem capacidade reduzidas, mesmo que geralmente substanciais, de guiar os usuários e uma percepção negativa quase imperceptível.

Os tipos de *Dark Pattern* também possuem variados níveis de influência, em alguns casos não se verificando efeitos significativos, conforme apurado no mesmo trabalho (LUGURI; STRAHILEVITZ, 2021). Resultados similares foram colhidos na pesquisa de Sin *et al.* (2022), os quais auferiram diferentes eficácias para cada *Dark Pattern* testado.

São circunstâncias impassíveis de serem ignoradas, pois a vedação ao uso de *Dark Patterns* em igual medida para hipóteses em que seus efeitos são substantivamente desproporcionais - por vezes indiscerníveis de uma interface usual - acarretaria a restrição injusta das atividades do sítio eletrônico. Isto segue a linha argumentativa de alguns comentários às Diretrizes 03/2022, que pugnaram pelo reconhecimento de uma escala gradativa para regulamentar os padrões (UNIÃO EUROPEIA, 2022).

Isto considerado, definir juridicamente os *Dark Patterns* em graus escalonados é uma tarefa árdua, tocando a celeuma da dificuldade em diferenciar *Dark Patterns* menos efetivos de uma prática legítima. Não se vislumbra outra escapatória a não ser incentivar a pesquisa empírica para averiguar quais tipos e quantidades de *Dark Pattern* produzem efeitos considerados substanciais e quais não os produzem, ao ponto de estes deixarem de ser considerados *Dark Patterns*.

Outra questão relevante trazida por Oliveira (2022) é a necessidade dos *Dark Patterns* serem tratados de acordo com o contexto das atividades do implementador, ocasionando que algumas interfaces condenadas pela doutrina possam gerar resultados benéficos dependendo de como forem utilizadas. V.g., a "Gamificação" (desbloqueio progressivo de funcionalidades com o uso da plataforma), descrita por Luguri e Strahilevitz (2021) como sendo um *Dark Pattern*, pode ser abusiva quando funções essenciais são bloqueadas por um número desbalanceado de usos ou investimentos, mas benéfica se inserida em plataformas ludificadas ou na introdução gradativa dos usuários a um aplicativo ou jogo de videogame. Da mesma forma, mensagens de estoque baixo, descritas por Mathur *et al.* (2019) como *Dark Patterns*, podem fazer sentido para informar um consumidor sobre os limites da capacidade de ofertar um serviço, se aplicadas comedidamente.

Uso interessante de interfaces tipicamente classificadas como *Dark Patterns* é para a manipulação dos usuários a tomar uma atitude saudável ou desejada, o que a doutrina denominou *Bright Patterns* (UNIÃO EUROPEIA, 2022), ou Padrões Claros. Hipoteticamente, por exemplo, um *Dark Pattern* do tipo "provocativo" poderia ser utilizado como *Bright Pattern* para repetidamente emitir um aviso sobre o tempo excessivo de uso de um aplicativo, com o intuito de zelar pela saúde mental de seus usuários. Coaduna-se com isto os achados do trabalho de Sin *et al* (2019), que apurou a efetividade de alguns métodos interventivos – interfaces que exercem influências reversas sobre o usuário – para compensar os efeitos de *Dark Patterns*.

Neste andar, a definição de *Dark Pattern* no CCPA/CPRA californiano, "interface de usuário desenhada e manipulada com o substancial efeito de subverter ou limitar a autonomia, liberdade de decidir ou de escolher do usuário, conforme definido em lei" (Cal.Civ.Code, §1798.140, l; tradução nossa), seria inadequada, pois desconsidera o uso de interfaces manipulativas para guiar o usuário em direção a práticas salutares.

Fora essas questões, os estudos jurídicos sobre os *Dark Patterns* são limitados quase exclusivamente ao direito do consumidor e da proteção de dados pessoais, quando a temática é pertinente a inúmeras outras áreas do direito. Ao direito de proteção à criança e ao adolescente e ao direito da pessoa idosa, v.g., importa compreender de que forma essas interfaces influenciam os sujeitos de direito tutelados, algo que infelizmente carece de estudos na doutrina. No mais, também se vislumbra espaço para analisar os *Dark Patterns* segundo a ótica do direito comercial e concorrencial, pois estas interfaces podem ser empregadas pelas empresas de forma abusiva durante o exercício de atividades comerciais; direito civil, no atinente à responsabilidade civil e a seara contratual; e do direito constitucional, quanto à proteção de direitos e garantias fundamentais dos afetados.

Esta interdisciplinaridade pode abarcar influências consideráveis de áreas além do direito, como a disciplina de UX/UI, do que se pode sublinhar os elementos do design ético citados por Bösch *et al* (2016)[103] – provavelmente úteis para direcionar a compreensão sobre quais interfaces podem ser aceitas ou não pelo ordenamento.

---

[103] Como citado no segundo capítulo, os elementos do design ético são: "proativo e não reativo; privacidade como configuração primária; privacidade como parte do design; funcionalidade plena; segurança *end-to-end* (do começo ao fim do tratamento de dados); visibilidade e transparência; e respeito à privacidade dos usuários" (BÖSCH *et al,* 2021; tradução nossa).

Isto conduz à noção de que, enquanto a proibição de uma prática potencialmente danosa aos usuários faz sentido, ela deve ser realizada levando em consideração os diferentes tipos de *Dark Patterns*, suas distintas capacidades, como são utilizados e as múltiplas facetas do direito nas quais interferem, bem como a possibilidade de novas espécies surgirem com o desenvolvimento tecnológico.

Assim, uma possível maneira de disciplinar os *Dark Patterns* no Brasil é através de um sistema composto de leis formais e regulamentos de órgãos temáticos da administração pública (PROCON, ANPD etc.) e autoridades setoriais, com o intuito de tecer uma malha normativa interdisciplinar que comporte alterações de maior agilidade, permitindo a célere complementação destas normas com os resultados de novos estudos empíricos e as peculiaridades de tecnologias emergentes.

Este modelo permite o controle gradual e cauteloso dos *Dark Patterns*, conciliando os esforços para estudá-los com a publicações de normas, diretrizes e guias orientativos sobre os conhecimentos sedimentados, e garante maior segurança aos fornecedores e agentes de tratamento de dados pessoais, que possuiriam instrumentos objetivos dos quais se valer para construir interfaces aceitas pelo direito.

## 5. CONCLUSÃO

Visando discutir a posição dos *Dark Patterns* no ordenamento jurídico brasileiro, foi possível, ao menos de maneira introdutória, estabelecer um conceito, situá-los quanto à sua potencial antijuridicidade e propor um modelo normativo para discipliná-los. Os resultados desta pesquisa pendem maiores desenvolvimentos acadêmicos, na medida que o tema ainda é trabalhado superficialmente no país. Sem embargos, a trajetória da pesquisa revelou muitos aspectos não previstos na formulação da hipótese primária.

No segundo capítulo, realizamos um breve apanhado histórico das tecnologias gráficas de interface humano-computador (*Human-Computer Interfaces*, ou HCI) até os atuais contornos das interfaces virtuais, inseridas nas disciplinas de UX (*User Experience* ou Experiência de Usuário) e UI (*User Interface* ou Interface de Usuário), com o intuito de destacar a importância do aperfeiçoamento destas ferramentas ao desenvolvimento e popularização da internet. Disso, abordamos a deturpação ética dos profissionais de UX/UI que permeia o desenvolvimento de interfaces, instigando a criação de designs capazes de influenciar consumidores a adotarem um comportamento que lhes é prejudicial, mas benéfico a quem o implementou – convencionando-se chamar essas técnicas de *Dark Patterns*, ou Padrões Obscuros.

Foi também evidenciado, a partir dos trabalhos do SERNAC (2021) e de Mathur *et al.* (2019), que a quantidade desses Padrões na web soma montas incalculáveis, apresentando-se em miríade de tipos e funções. Desta pluralidade de formatos e da novidade da temática, que passou a ser abordada principalmente a partir da iniciativa de Harry Brignull em 2010, decorrem a ausência de concordância sobre seus conceitos e características pela doutrina.

Apesar de existir número relativamente baixo de trabalhos sobre os efeitos dos *Dark Patterns*, dois experimentos apresentados no célebre artigo de Luguri e Strahilevitz (2021) apresentaram fortes comprovações de suas eficácias. A primeira pesquisa mostrou que a permanência no serviço e os impactos psicológicos aos usuários aumentaram proporcionalmente conforme a intensidade dos *Dark Patterns* empregados, sem que houvesse um impacto significativo dos preços do serviço ofertado. As variáveis demográficas dos participantes do experimento apontaram que pessoas com menor grau de escolaridade foram mais suscetíveis aos *Dark Patterns.*

Na segunda pesquisa auferiu-se diferentes níveis de eficácia para cada tipo de *Dark Pattern*, ao passo que os resultados do primeiro estudo foram novamente replicados. (LUGURI; STRAHILEVITZ, 2021) Essas conclusões foram também encontradas nos trabalhos de outros pesquisadores, como Sin *et al.* (2019) e Blake *et al.* (2018).

Assim, acreditamos que o segundo capítulo conseguiu demonstrar um considerável interesse jurídico no estudo da temática, mas em moldes imprevistos à formulação da hipótese principal. Apesar da existência de estudos metodologicamente rigorosos que demonstraram as suas proporções e os enormes efeitos que causam nas decisões do consumidor - mesmo que em graus variados, de acordo com diferentes espécies de Padrões -, temos poucos conhecimentos de como os mecanismos e dinâmicas atreladas a essas técnicas se interrelacionam e manifestam. A matéria ainda foi pouco esculpida pela doutrina, situando-nos frente a uma incompreendida temática de singular interesse social.

Para averiguar como os *Dark Patterns* estão sendo abordados globalmente, no terceiro capítulo foram estudados posicionamentos, casos e normas jurídicas relevantes nos EUA, União Europeia, e OCDE.

Nos Estados Unidos, a atuação da Comissão Federal de Comércio (FTC) contra os *Dark Patterns* é a que mais se destaca, enfrentando-os rigorosamente através de litígios e da elaboração de pareceres e relatórios, balizando a disciplina da temática no país. Em todas as suas manifestações a FTC se posicionou de forma bastante dura frente aos *Dark Patterns*, não tolerando o seu uso em atividades comerciais.

Os esboços das leis federais "Ato para a Redução de Experiências Enganosas a Usuários Online" (*Deceptive Experiences To Online Users Reduction Act*, ou DETOUR) e "Ato sobre Dados Seguros" (*Safe Data Act*) também combatem frontalmente o uso *Dark Patterns*, no tocante às atividades de fornecedores online e à proteção de dados pessoais. Similar entendimento foi consolidado nas leis de proteção de dados dos consumidores da Califórnia e do Colorado (CCPA/CPRA e CPA), que proibiram o uso de *Dark Patterns* para a coleta do consentimento, além de argumentarmos pela lesividade destas técnicas a normas gerais de proteção de dados de modelo europeu (DONEDA, 2021), contemplando o VCDPA do Estado da Virgínia.

Na União Europeia, o Comitê de Proteção de Dados Europeu (EDPB) editou as Diretrizes 03/2022 sobre *Dark Patterns* nas redes sociais (UNIÃO EUROPEIA, 2022), através do que se posicionou de forma contrária a essas técnicas. No documento, foram conceituadas várias práticas distintas que configurariam *Dark Patterns*, relacionando-as com violações a dispositivos do Regulamento Geral de Proteção de Dados (RGPD).

Diversas entidades e indivíduos de todo o mundo se manifestaram sobre as Diretrizes 03/2022, levantando sugestões, críticas e elogios à EDPB. Em geral, os comentários reforçaram a inexistência de uma unicidade teórica sobre como regular os *Dark Patterns*, trazendo múltiplas visões distintas da comunidade acadêmica que atestam a complexidade do tema. Entre alguns dos tópicos levantados, é possível destacar a sugestão de reconhecer a vulnerabilidade dos usuários da internet; a necessidade das normas jurídicas de levar em consideração a intensidade dos *Dark Patterns* e seus tipos; a existência de interfaces manipulativas que possuem bons propósitos - os *Bright Patterns*, ou Padrões Claros -; e a dificuldade de distinguir algumas práticas aceitáveis de *Dark Patterns*.

Por fim, o posicionamento da OCDE sobre os *Dark Patterns* se mostra importante não apenas por sua precisão técnica, mas também ao evidenciar que a matéria alcança um patamar de discussão elevado – chegando a entidade internacional de profunda pertinência na moderna conjuntura político-econômica.

A entidade constatou as mesmas dificuldades conceituais sentidas ao início da pesquisa, citando o trabalho de Mathur *et al*. (2021) Sobre essas imprecisões. Nesta senda, apesar de reconhecer sua potencialidade danosa, a OCDE reafirmou a dificuldade casuística em diferenciar interfaces comuns de *Dark Patterns* e, com isso, de encontrar a melhor forma de regulá-los, ressaltando o papel das instituições especializadas em defesa do consumidor de fomentar a pesquisa sobre estas tecnologias.

Em suma, o estudo sobre as experiências estrangeiras com os *Dark Patterns* reforçou o constatado pela doutrina apurada no segundo capítulo: apesar da indeterminação conceitual sobre os *Dark Patterns*, o que é espelhado nos comentários diversificados às Diretrizes 03/2022 da EDPB, a matéria deve ser vista com relevância, resultando em um movimento inicial para a proibição destas técnicas. Uma maior unidade de tratamento aos *Dark Patterns* foi vista nos EUA, mas as discussões havidas no país, na União Europeia e na OCDE demonstram que se está longe de um consenso sobre como equacioná-los.

No quarto capítulo partiu-se à análise gradual, em etapas, dos *Dark Patterns* e sua regulamentação no ordenamento jurídico brasileiro, utilizando como alicerce teórico o direito nacional prévio em comparação com o conteúdo levantado nos capítulos 2 e 3.

A começar pela análise da doutrina brasileira, averiguou-se que ela ainda é bastante imatura, encontrando-se total de 5 estudos metodologicamente coesos que a abordam segundo as perspectivas do direito, administração e, destacadamente, das ciências da computação. Em grande parte repisando os achados de outros estudos estrangeiros, essas pesquisas trouxeram alguns apontamentos de relevância para o desenvolvimento da temática no Brasil, dos quais sublinhamos a violação dos *Dark Patterns* à Lei Geral de Proteção de Dados Pessoais (LGPD) e alguns aprofundamentos conceituais, mormente a introdução do termo *Dark Strategy* (CARNEIRO, 2022) e a problemática da intencionalidade dos *Dark Patterns*, elemento que frequentemente integra suas definições apesar de ser dificilmente apurado e acarretar a desconsideração de interfaces que acidentalmente causam danos (OLIVEIRA, 2022).

Compreendida a situação acadêmica dos *Dark Patterns* no Brasil, iniciamos tentando vencer o problema conceitual que há muito circunda o tema. Por isso, com o intuito de averiguar pontos de contato comuns ou a existência de um conceito capaz de abarcar as principais características do fenômeno, tabelamos 16 nomenclaturas e conceitos usados na doutrina internacional e em normas jurídicas.

A partir desse arranjo, encontramos as mesmas observações feitas no trabalho de Mathur *et al*. (2021), que realizaram pesquisa similar ressaltando diversos pontos de dissonância sobre características e qualidades atribuídas aos *Dark Patterns* pela doutrina. Somado a isso, destacamos que a intencionalidade na aplicação dos *Dark Patterns* se fez presente na maioria dos conceitos, como avaliado por Jônatas Kerr de Oliveira (2022).

Considerando a tutela dos indivíduos vulneráveis como os principais objetivos da proteção de dados pessoais e do direito do consumidor, tecemos a seguinte definição jurídica para o objeto: *Dark Pattern* consiste em toda interface capaz de substantivamente violar direitos e causar danos à pessoa natural. Trata-se de um conceito prototípico consideravelmente discrepante com os demais propostos em estudos e normas jurídicas, mas que sem embargos pode ser adequado para proteger os consumidores dos danos causados por essas interfaces.

Possuindo em mãos este conceito, analisamos os *Dark Patterns* sob a perspectiva do direito do consumidor e da proteção de dados, a partir dos quais argumentamos que essas técnicas podem violar a ordem jurídica, malferindo princípios, regras da oferta, requisitos para o consentimento válido e a escolha de bases legais para o tratamento, entre outras disposições, nos casos em que seus efeitos forem comprovadamente negativos aos sujeitos de direito.

Entretanto, concluiu-se que a declaração de sua antijuridicidade, apesar de evidente em alguns casos, deve ser vista com cautela. Quando os *Dark Patterns* são utilizados de forma desmedida a antijuridicidade é identificável facilmente, mas, considerando que interfaces aplicadas de forma menos "intensa" podem possuir efeitos reduzidos, bem como as diferentes espécies de *Dark Patterns* apresentam níveis distintos de eficácia (LUGURI; STRAHILEVITZ, 2021), a discriminação entre uma interface jurídica e antijurídica não é sempre uma tarefa simples. Também merece atenção o uso de interfaces manipulativas para fins éticos, na forma de *Bright Patterns*, de maneira que a proibição integral desse tipo de técnicas não parece ser a melhor solução para todas as hipóteses.

Por isso, sugerimos que a regulamentação dos *Dark Patterns* no Brasil ocorra por intermédio da cooperação entre o Poder Legislativo, entidades especializadas da administração pública e órgãos setoriais, com o intuito de estabelecer leis, regulamentos, diretrizes e orientações para abordar a temática de maneira abrangente e interdisciplinar, devendo ser incentivados novos estudos sobre essas técnicas.

Tendo em vista essas considerações, a hipótese deste estudo, que os *Dark Patterns* devem ser incondicionalmente declarados antijurídicos e banidos do ordenamento, não pode ser atualmente sustentada. Apesar das graves evidências sobre o potencial destas técnicas para manipular, enganar e confundir os usuários, a complexidade do tema acarreta que a regulamentação dos *Dark Patterns* pelo ordenamento jurídico brasileiro não pode se bastar em proibi-los, sendo necessário estabelecer orientações e normas que abarquem, além de mecanismos sancionatórios, quais interfaces são aceitáveis ou inaceitáveis, tendo em vista as múltiplas áreas do direito às quais interessa a regulamentação dos *Dark Patterns* e a facilitação da conformidade para agentes de tratamento e fornecedores.

Em arremate, as pesquisas sobre os *Dark Patterns* são território fértil para o direito, que ainda busca as respostas corretas sobre como moderá-los apropriadamente. Espera-se, portanto, que este trabalho seja um estopim, ou ao menos uma fagulha, para captar o interesse da academia pelos *Dark Patterns*, com os votos de que possamos iluminar o caminho em direção ao equilíbrio nas relações digitais.

# REFERÊNCIAS

ANDRADE, Landolfo; MAGRO, Américo Ribeiro. **Manual de Direito Digital**. São Paulo: Jus Podivm, 2021.

BARNES, Susan B. **User Friendly**: A Short History of the Graphical User Interface. Sacred Heart University Review. Vol. 16, n.º 1, 2010. Disponível em: https://digitalcommons.sacredheart.edu/shureview/vol16/iss1/4/?utm_source=digitalc ommons.sacredheart.edu%2Fshureview%2Fvol16%2Fiss1%2F4&utm_medium=PD F&utm_campaign=PDFCoverPages. Acesso em: 04 out 2021.

BESEMER, Leo. **Privacy and Data Protection Based on the GDPR**: understanding the general data protection regulation. Amersfoort: Van Haren Publishing, 2020.

BEUC. **"Dark Patterns" and the EU Consumer Law Acquis**: Recommendations for better enforcement and reform. 2022. Disponível em: https://www.beuc.eu/sites/default/files/publications/beuc-x-2022-013_dark_patters_paper.pdf. Acesso em: 03 out 2022.

BIONI, Bruno Ricardo. **Proteção de Dados Pessoais**: a função e os limites do consentimento. 3ª ed. rev., atual. e ampl. Rio de Janeiro: Forense, 2021.

BIONI, Bruno Ricardo; LUCIANO, Maria. O Consentimento como Processo: em Busco do Consentimento Válido. *In:* MENDES, Laura Schertel *et al* (coord.). **Tratado de Proteção de Dados Pessoais**. Rio de Janeiro: Forense, 2021. P. 149-161.

BLAKE, Thomas; *et al.* **Price Salience and Product Choice**. Cambridge: National Bureau of Economic Research, 2018. Disponível em: https://www.nber.org/system/files/working_papers/w25186/w25186.pdf. Acesso em: 11 set 2022.

BOBBIO, Norberto. **Teoria Geral do Direito**. 3ª ed. São Paulo: Martins Fontes, 2010.

BÖSCH, Christoph; ERB, Benjamin; KARGL, Frank; KOPP, Henning; PFATTHEICHER, Stefan. Tales from the Dark Side: Privacy Dark Strategies and Privacy Dark Patterns. **Proceedings on Privacy Enhancing Technologies**, julho, 2016, p. 237-254. Disponível em: https://doi.org/10.1515/popets-2016-0038. Acesso em 21 set 2021.

BRASIL. **Lei n.º 13.709, de 14 de agosto de 2018**. Lei Geral de Proteção de Dados Pessoais (LGPD). Disponível em: http://www.planalto.gov.br/ccivil_03/_ato2015-2018/2018/lei/l13709.htm. Acesso em: 16 out 2022.

BRASIL. **Lei n.º 8.078, de 11 de setembro de 1990**. Dispõe sobre a proteção do consumidor e dá outras providências. Disponível em: http://www.planalto.gov.br/ccivil_03/leis/l8078compilado.htm. Acesso em: 21 out 2022.

BRIGNULL, Harry. **Dark Patterns**: dirty tricks designers use to make people do stuff. 90percentofeverything, 2010. Disponível em: https://90percentofeverything.com/2010/07/08/dark-patterns-dirty-tricks-designers-use-to-make-people-do-stuff/index.html. Acesso em: 16 set 2022.

BRIGNULL, Harry. **Dark Patterns**: inside the interfaces designed to trick you. TheVerge, 2013. Disponível em: https://www.theverge.com/2013/8/29/4640308/dark-patterns-inside-the-interfaces-designed-to-trick-you. Acesso em: 16 out 2022.

CALIFÓRNIA. **California Civil Code, §§1798.99.28-1798.99.32**. California Age-Appropriate Design Code Act. Disponível em: https://leginfo.legislature.ca.gov/faces/billTextClient.xhtml?bill_id=202120220AB2273. Acesso em: 17 out 2022.

CALIFÓRNIA. **California Civil Code, §§1798.100-1798.199.100**. California Consumer Privacy Act of 2018. Disponível em: https://leginfo.legislature.ca.gov/faces/codes_displayText.xhtml?lawCode=CIV&division=3.&title=1.81.5.&part=4.&chapter=&article=. Acesso em: 17 out 2022.

CALONGA, Luiz Octavio Lanssoni *et al*. Pensa que me Engana, eu Finjo que Acredito: Padrões Obscuros sob a Perspectiva do Usuário. **XLVI Encontro da ANPAD - EnANPAD 2022**. set. 2022. Disponível em: http://anpad.com.br/uploads/articles/120/approved/8a88d5f412f2ad376f8597d28cbd3720.pdf. Acesso em: 16 out 2022.

CARNEIRO, Eduardo Teixeira. **Dark Patterns na Comunicação de Cookies em Sites de Notícias**. Trabalho de Conclusão de Curso (Graduação) - Universidade Federal Fluminense, Niterói, 2022. Disponível em: https://app.uff.br/riuff/bitstream/handle/1/25769/EDUARDO%20TEIXEIRA%20CARNEIRODARK%20PATTERNS%20NA%20COMUNICA%C3%87%C3%83O%20DE%20COOKIES%20EM%20SITES%20DE%20NOT%C3%8DCIAS%20-%20VERS%C3%83O%20FINAL.pdf?sequence=1. Acesso em: 16 out 2022.

CAVOUKIAN, Ann; *et al*. **Privacy by design**: The 7 foundational principles. Canada: 2009. Disponível em: http://jpaulgibson.synology.me/ETHICS4EU-Brick-SmartPills-TeacherWebSite/SecondaryMaterial/pdfs/CavoukianETAL09.pdf. Acesso em: 27 set 2022.

CHOPRA, Rohit. **Statement Of Comissioner Rohit Copra Regarding Dark Patterns in the Matter of Age of Learning, Inc. Federal Trade Comission**. 2020. Disponível em: https://www.ftc.gov/system/files/documents/public_statements/1579927/172_3086_abcmouse_-_rchopra_statement.pdf. Acesso em: 08 set 2022.

COHEN-ALMAGOR, Raphael. Internet History. **International Journal of Technoethics**. Vol. 2, n.º 2, Abril-junho 2011, p. 45-64. Disponível em: http://www.sites.upiicsa.ipn.mx/archivos/profesores/jlopez/2019-2020-1/web/presentaciones/Internet_History.pdf. Acesso em: 10 set 2022.

COLORADO. **Colorado Revised Statutes, §§6-1-1301-6-1-1313**. Colorado Privacy Act. Disponível em:
https://advance.lexis.com/documentpage/?pdmfid=1000516&crid=59ea1035-8a6c-486f-94fc-deafaa8df1ad&nodeid=AAGAABAABAAOAAB&nodepath=%2FROOT%2FAAG%2F
AAGAAB%2FAAGAABAAB%2FAAGAABAABAAO%2FAAGAABAABAAOAAB&level
=5&haschildren=&populated=false&title=Part+13+Colorado+Privacy+Act+(Effective+
July+1%2C+2023)&config=014FJAAyNGJkY2Y4Zi1mNjgyLTRkN2YtYmE4OS03NT
YzNzYzOTg0OGEKAFBvZENhdGFsb2d592qv2Kywlf8caKqYROP5&pddocfullpath=
%2Fshared%2Fdocument%2Fstatutes-legislation%2Furn%3AcontentItem%3A63JP-
6XP3-GXJ9-330K-00008-00&ecomp=8gf59kk&prid=ed256438-0a5b-4f48-8452-
cfe69e1ab945. Acesso em: 17 out 2022.

CONTI, Gregory; SOBIESK, Edward. Malicious Interface Design: Exploiting the User. In: **Proceedings of the 19th international conference on World wide web**. 2010. p. 271-280. Disponível em:
https://dl.acm.org/doi/abs/10.1145/1772690.1772719?casa_token=igblAfKqUsoAAA
AA:n1KYg4rC7IaBs9_AsGOYRLTaFrqtGnLK2SdcQ5cuDoDNXd1s4HJp-
X1jxc7kcUq_KUPfSrc5Kl_ktw. Acesso em: 10 set 2022.

CYBER.CL. **Sobre Cyber**. Disponível em:
https://www.cyber.cl/index.html#information/preguntas-frecuentes. Acesso em: 11 set 2022.

DHENAKARAN, S.S.; SAMBANTHAN, K. Thirugnana. Web Crawler - an Overview. **International Journal of Computer Science and Communication**. Vol. 2, No. 1, 2011, p. 265-267. Disponível em: http://www.csjournals.com/IJCSC/PDF2-1/Article_49.pdf. Acesso em: 07 out 2022.

DONEDA, Danilo. **Da Privacidade à Proteção de Dados Pessoais**: fundamentos da ILei Geral de Proteção de Dados Pessoais. 3ª ed. São Paulo: Thomson Reuters Brasil, 2022.

ESTADOS UNIDOS DA AMÉRICA. UNITED STATES DISTRICT COURT CENTRAL DISTRICT OF CALIFORNIA. **Case No. 2:20-cv-7996**. 10 ago 2020. Disponível em: https://www.ftc.gov/legal-library/browse/cases-proceedings/172-3186-age-learning-inc-abcmouse. Acesso em: 14 out 2022.

ESTADOS UNIDOS DA AMÉRICA. **16 CFR Part 310**. 1995. Telemarketing Sales Rules. Disponível em: https://www.ftc.gov/legal-library/browse/rules/telemarketing-sales-rule. Acesso em: 08 set. 2022.

ESTADOS UNIDOS DA AMÉRICA. **16 CFR Part 316**. 2003. CAN-SPAM Rule. Disponível em: https://www.ftc.gov/legal-library/browse/rules/can-spam-rule. Acesso em: 17 out 2022.

ESTADOS UNIDOS DA AMÉRICA. **S.2499-117th Congress**. Setting an American Framework to Ensure Data Access, Transparency, and Accountability Act or the SAFE DATA Act. Disponível em: https://www.congress.gov/bill/117th-congress/senate-bill/2499. Acesso em: 17 out 2022.

ESTADOS UNIDOS DA AMÉRICA. **S.3330-117h Congress**. Deceptive Experiences To Online Users Reduction Act or the DETOUR Act. Disponível em: https://www.congress.gov/bill/117th-congress/senate-bill/3330. Acesso em: 17 out 2022.

ESTADOS UNIDOS DA AMÉRICA. **15 U.S.C. §§ 41-58**. Federal Trade Comission Act. 1914. Disponível em: https://www.ftc.gov/legal-library/browse/statutes/federal-trade-commission-act. Acesso em: 08 set. 2022

ESTADOS UNIDOS DA AMÉRICA. **15 U.S.C. §§ 8401-8405**. Restore Online Shoppers Confidence Act. 2010. Disponível em: https://www.ftc.gov/legal-library/browse/statutes/restore-online-shoppers-confidence-act. Acesso em: 08 set. 2022.

FAZLIOGLU, Müge. **Consolidating US privacy legislation**: The SAFE DATA Act. IAPP, 2020. Disponível em: https://iapp.org/news/a/consolidating-u-s-privacy-legislation-the-safe-data-act/. Acesso em: 19 set. 2022

FEDERAL TRADE COMMISSION. **Bringing Dark Patterns to Light Staff Report**. 2022. Disponível em: https://www.ftc.gov/reports/bringing-dark-patterns-light. Acesso em 20 set. 2022.

FEDERAL TRADE COMMISSION. **Children's Online Learning Program ABCmouse to Pay $10 Million to Settle FTC Charges of Illegal Marketing and Billing Practices**. 2020. Disponível em: https://www.ftc.gov/news-events/news/press-releases/2020/09/childrens-online-learning-program-abcmouse-pay-10-million-settle-ftc-charges-illegal-marketing. Acesso em: 08 set. 2022.

FEDERAL TRADE COMISSION. **Combatting Online Harms Through Innovation**: A Report to Congress. 2022. Disponível em: https://www.ftc.gov/system/files/ftc_gov/pdf/Combatting%20Online%20Harms%20Through%20Innovation%3B%20Federal%20Trade%20Commission%20Report%20to%20Congress.pdf. Acesso em: 08 set. 2022.

FEDERAL TRADE COMMISSION. **FILE NO. 202 3138**. 1 set 2022. Disponível em: https://www.ftc.gov/legal-library/browse/cases-proceedings/2023138-credit-karma-llc. Acesso em: 15 out 2022.

GRAY, Colin M.; KOU, Yubo; BATTLES, Bryan; HOGGATT, Joseph; TOOMBS, Austin L. The Dark (Patterns) Side of UX Design. **CHI '18**: Proceedings of the 2018 CHI Conference on Human Factors in Computing Systems. Paper n.º 534, Abril 2018, P. 1-14. Disponível em: https://doi.org/10.1145/3173574.3174108. Acesso em 20 set 2021.

GREVE, Bent. What is Welfare? **Central European Journal of Public Policy**. Vol. 2, n.º 1, 2008, p. 50-73. Disponível em: https://www.researchgate.net/profile/Bent-Greve/publication/26591601_What_is_Welfare/links/0f31753c4fa4faac80000000/What-is-Welfare.pdf. Acesso em: 03 out 2022.

INTERFACE. *In:* **Dicionário Priberam da Língua Portuguesa**. 2021. Disponível em. https://dicionario.priberam.org/interface. Acesso em: 19 out 2022.

JOO, Heonsik. A Study on Understanding of UI and UX, and Understanding of Design According to User Interface Change. **International Journal of Applied Engineering Research**. Vol. 12, n.º 20, 2017, p. 9931-9935. Disponível em: http://www.ripublication.com/ijaer17/ijaerv12n20_96.pdf. Acesso em: 06 out 2021.

JÚNIOR, Humberto Theodoro. **Direitos do Consumidor**. 10ª ed. Rev., atual. e ampli. Rio de Janeiro: Forense, 2021.

LEMOS, André; MARQUES, Daniel. Interfaces Maliciosas: estratégias de coleta de dados pessoais em aplicativos. **V!RUS**. São Carlos, n.º 19, 2019. Disponível em: http://www.nomads.usp.br/virus/virus19/?sec=4&item=2&lang=pt. Acesso em: 07 Out. 2022.

LIMA, Patrícia Raposo Santana. **Investigação da Comunicabilidade e Uso de Dark Patterns em Privacidade no Instagram**. Trabalho de Conclusão de Curso (Graduação) - Universidade Federal Fluminense, Niterói, 2021. Disponível em: https://app.uff.br/riuff/bitstream/handle/1/25726/PATRICIA%20RAPOSO%20TCC%20MonografiaPatriciaRaposoVersaoFinal.pdf?sequence=1. Acesso em: 16 out 2022.

LUGURI, Jamie; STRAHILEVITZ, Lior Jacob. Shining a Light on Dark Patterns. **Journal of Legal Analysis**. Vol. 13, n.º 1, 2021, p. 43-109. Disponível em: https://academic.oup.com/jla/article/13/1/43/6180579. Acesso em: 14 nov 2021.

MACKAY, Wendy E.; FAYARD, Anne-Laure. HCI, Natural Science and Design: A Framework for Triangulation Across Disciplines. **DIS'97**: Designing Interactive Systems. Agosto, 1997. P. 223-234. Disponível em: https://doi.org/10.1145/263552.263612. Acesso em: 15 set 2021.

MARCULINO, Karoline Silveira. **Vulnerabilidade do Titular de Dados Pessoais e a Responsabilidade dos Agentes de Tratamento**. Trabalho de Conclusão de Curso (graduação) – Universidade Federal do Rio Grande do Sul, Porto Alegre, 2021. Disponível em: https://www.lume.ufrgs.br/bitstream/handle/10183/237640/001139342.pdf?sequence=1. Acesso em: 26 out 2021.

MATHUR, Arunesh; *et al.* Dark Patterns at Scale: Findings from a Crawl of 11K Shopping Websites. **Proceedings of the ACM on Human-Computer Interactions**. Vol. 3, N.º CSCW, artigo 81, 2019. Disponível em: https://arxiv.org/pdf/1907.07032.pdf?utm_source=https://humansplustech.substack.com. Acesso em: 20 out 2021.

MATHUR, Arunesh; *et al.* What Makes a Dark Pattern… Dark? Design attributes, normative considerations, and measurement methods. **Proceedings of the 2021 CHI conference on human factors in computing systems**. 2021. p. 1-18. Disponível em: https://dl.acm.org/doi/pdf/10.1145/3411764.3445610?casa_token=ae8x4c2oz6YAAAAA:VfVrKr8fK9n1rhl-

tjG2r4_njfo9gu4PAjinsRwut78DySWx4JvYp_GI_rs9sjc93T0C2PQY-fOHuA0. Acesso em: 14 set 2022.

NUNES, Rizzatto. **Curso de Direito do Consumidor**. 14ª ed. São Paulo: Saraiva, 2021.

OLIVEIRA, Jônatas Kerr de. **Coerção, Manipulação e Tecnologia**: Estratégias de Monetização em Jogos *Free-to-Play*. Dissertação (doutorado) - Universidade Federal de São Carlos, São Carlos, 2022. Disponível em: https://repositorio.ufscar.br/bitstream/handle/ufscar/16480/Tese_JKO_Manipulacao_ Coercao_Tecnologia_Final_Folha_Aprovacao.pdf?sequence=1. Acesso em: 16 out 2022.

PALHARES, Felipe. *Cookies*: contornos atuais. *In:* PALHARES, Felip (org.) *et al.* **Temas Atuais de Proteção de Dados**. 2ª ed. rev. e amp. São Paulo: Thomson Reuters Brasil, 2022. P. 11-61.

REIMER, Jeremy. A History of the GUI. **Ars Technica**, v. 5, p. 1-17, 2005. Disponível em: https://www.readit-dtp.de/PDF/gui_history.pdf. Acesso em: 06 out 2021.

RICART, Glenn. The Mosaic Web Browser. **Computer in Physics**. Vol. 8, 1994, p. 249-253. Disponível em: https://doi.org/10.1063/1.4823294. Acesso em: 29 set 2021.

SERNAC. **Informe de Resultados de Levantamiento de Dark Patterns en Comercio Electrónico**. Chile: 2021. Disponível em: https://www.sernac.cl/portal/617/articles-62983_archivo_01.pdf. Acesso em 20 set 2021.

SIN, Ray *et al.* Dark patterns in online shopping: do they work and can nudges help mitigate impulse buying?. In: **Behavioural Public Policy**. 2022, p. 1-27. Disponível em: https://doi.org/10.1017/bpp.2022.11. Acesso em: 30 jun 2022.

SOLER, Sebastián. **Ley, Historia y Libertad**. Argentina: Olejnik, 2019.

STANFORD UNIVERSITY. **Artificial Intelligence Index Report 2022**. 2022. Disponível em: https://aiindex.stanford.edu/wp-content/uploads/2022/03/2022-AI-Index-Report_Master.pdf. Acesso em 29 out 2022.

TARTUCE, Flávio. **Manual de Direito Civil: volume único**. 11ª ed. rev. atual. e ampli. Rio de Janeiro: Forense; Método, 2021.

UNIÃO EUROPEIA. **Guia Prático Comum do Parlamento Europeu, do Conselho e da Comissão para as Pessoas que Contribuem para a Redação de Textos Legislativos da União Europeia**. Luxemburgo: Serviço das Publicações da União Europeia, 2016. Disponível em: https://eur-lex.europa.eu/content/techleg/PT-guia-de-redacao-de-textos-legislativos.pdf. Acesso em: 26 set 2022.

UNIÃO EUROPEIA. **Guidelines 3/2022 on Dark patterns in social media platform interfaces**: How to recognise and avoid them. 2022. Disponível em: https://edpb.europa.eu/system/files/2022-03/edpb_03-2022_guidelines_on_dark_patterns_in_social_media_platform_interfaces_en.pdf. Acesso em: 10 set 2022.

UNIÃO EUROPEIA. Guidelines 03/2022 on Dark patterns in social media platform interfaces: How to recognise and avoid them. 2022. **Feedback**. Disponível em: https://edpb.europa.eu/our-work-tools/documents/public-consultations/2022/guidelines-32022-dark-patterns-social-media_en. Acesso em: 14 set 2022.

UNIÃO EUROPEIA. **Position of the European Parliament adopted at first reading on 5 July 2022 with a view to the adoption of Regulation (EU) 2022/… of the European Parliament and of the Council on a Single Market For Digital Services (Digital Services Act) and amending Directive 2000/31/EC**. 2022. Disponível em: https://www.europarl.europa.eu/doceo/document/TA-9-2022-0269_EN.html. Acesso em: 03 out 2022.

UNIÃO EUROPEIA. **Proposal for a REGULATION OF THE EUROPEAN PARLIAMENT AND OF THE COUNCIL LAYING DOWN HARMONISED RULES ON ARTIFICIAL INTELLIGENCE (ARTIFICIAL INTELLIGENCE ACT) AND AMENDING CERTAIN UNION LEGISLATIVE ACTS**. 2021. Disponível em: https://eur-lex.europa.eu/legal-content/EN/TXT/?uri=CELEX:52021PC0206. Acesso em: 29 out 2022.

VIRGINIA. **Code of Virginia, §§59.1-575-59.1-585**. Consumer Data Protection Act. Disponível em: https://law.lis.virginia.gov/vacode/title59.1/chapter53/. Acesso em: 17 out 2022.

ZAGAL, José P.; BJÖRK, Staffan; LEWIS, Chris. Dark Patterns in the Design of Games. In: **Foundations of Digital Games Conference**, FDG 2013, Maio 2013, Chania, Grécia. Disponível em: https://my.eng.utah.edu/~zagal/Papers/Zagal_et_al_DarkPatterns.pdf. Acesso em 20 set 2021.